\documentclass[traditabstract]{aa} 
\usepackage{graphicx}
\usepackage{txfonts}
\usepackage{lscape}
\usepackage[figuresright]{rotating}
\usepackage{array}
\usepackage{dcolumn}
\usepackage{amssymb}
\usepackage[round]{natbib}
\bibpunct{(}{)}{;}{a}{}{,}
\def\ion#1#2{{\rm #1}\,{\sc #2}}
\def\vexp{\ifmmode \upsilon_{\rm exp} \else $\upsilon_{\rm exp}$\fi}
\def\taua{\ifmmode \tau_{a}\else $\tau_{a}$\fi}
\begin{document}
\title{Rest-frame ultraviolet spectrum of the gravitationally lensed galaxy 
`the 8~o'clock arc': stellar and interstellar medium properties
\thanks{Based on X-shooter observations made with the European Southern
Observatory VLT/Melipal telescope, Paranal, Chile, collected during the first
X-shooter Commissioning run.}}


\author{M. Dessauges-Zavadsky\inst{1}, 
        S. D'Odorico\inst{2},
	D. Schaerer\inst{1,3},
	A. Modigliani\inst{2},
	C. Tapken\inst{4,5},
	\and
	J. Vernet\inst{2}
       }

\offprints{miroslava.dessauges@unige.ch}

\institute{Observatoire de Gen\`eve, Universit\'e de Gen\`eve, 51 Ch. des
           Maillettes, 1290 Sauverny, Switzerland
          \and
           European Southern Observatory, Karl-Schwarzschildstr. 2, 85748
	   Garching bei M\"unchen, Germany
	   \and
	   Laboratoire d'Astrophysique de Toulouse-Tarbes, Universit\'e de 
	   Toulouse, CNRS, 14 Avenue E. Belin, 31400 Toulouse, France
	   \and
	   Max-Planck-Institut f\"ur Astronomie, K\"onigstuhl 17, 69117
	   Heidelberg, Germany
	   \and
	   Astrophysikalisches Institut Potsdam, An der Sternwarte 16, 14482
	   Potsdam, Germany
          }

\date{}

\authorrunning{M. Dessauges-Zavadsky et al.}

\titlerunning{Rest-frame UV spectrum of the 8~o'clock arc}

\abstract{We present the first detailed analysis of the rest-frame ultraviolet 
spectrum of the gravitationally lensed Lyman break galaxy (LBG), the 
``8~o'clock arc'', obtained with the intermediate-resolution X-shooter 
spectrograph recently commissioned on the ESO Very Large Telescope. Besides 
MS\,1512--cB58, the Cosmic Horseshoe, and the Cosmic Eye, three other lensed 
LBGs at comparable redshifts, this is the fourth of such a study, usually 
unfeasible at high redshifts. The spectrum of the 8~o'clock arc is rich in 
stellar and interstellar features, and presents several similarities to the 
well-known MS\,1512--cB58 LBG. The stellar photospheric absorption lines 
allowed us to constrain the systemic redshift, $z_{\rm sys} = 2.7350\pm 
0.0003$, of the galaxy, and derive its stellar metallicity, $Z=0.82~Z_{\odot}$, 
which is in excellent agreement with the metallicity determined from nebular 
emission lines. With a total stellar mass of $\sim 4.2\times 
10^{11}$~M$_{\odot}$, the 8~o'clock arc agrees with the mass-metallicity 
relation found for $z>2$ star-forming galaxies, although being located near the
upper end of the distribution given its high mass and high metallicity. Broad 
\ion{He}{ii}\,$\lambda$1640 emission is found, indicative of the presence of 
Wolf-Rayet stars formed in an intense period of star formation. The 31 
interstellar absorption lines detected led to the abundance measurements of 9 
elements. The metallicity of the interstellar medium (ISM), $Z=0.65~Z_{\odot}$ 
(Si), is very comparable to the metallicity of stars and ionized gas, and 
suggests that the ISM of the 8~o'clock arc has been rapidly polluted and 
enriched by ejecta of OB stars. The ISM lines extend over a very large velocity 
range, $\Delta \upsilon \sim 1000$ km~s$^{-1}$, from about $-800$ to $+300$ 
km~s$^{-1}$ relative to the systemic redshift, and have their peak optical 
depth blueshifted relative to the stars, implying gas outflows of 
$\upsilon_{\rm ISM} \simeq -120$ km~s$^{-1}$. The zero residual intensity in 
the strongest lines indicates a nearly complete coverage of the UV continuum by 
the ISM. The Ly$\alpha$ line is dominated by a damped absorption profile on top 
of which is superposed a weak emission, redshifted relative to the ISM lines by 
about +690 km~s$^{-1}$ and resulting from multiply backscattered Ly$\alpha$ 
photons emitted in the \ion{H}{ii} region surrounded by the cold, expanding ISM 
shell. A homogeneous spherical shell model with a constant outflow velocity, 
determined by the observations, is able to reproduce the observed Ly$\alpha$ 
line profile. Furthermore, the required dust content, $E(B-V) \approx 0.3$, is
in good agreement with the attenuation measured from the Balmer decrement. 
These results obtained from the radiation transfer modeling of the Ly$\alpha$ 
line in the 8~o'clock arc fully support the scenario proposed earlier, where 
the diversity of Ly$\alpha$ line profiles in Lyman break galaxies and 
Ly$\alpha$ emitters, from absorption to emission, is mostly due to variations 
of \ion{H}{i} column density and dust content.}

\keywords{cosmology: observations -- galaxies: individual: 8~o'clock arc --
galaxies: starburst -- galaxies: abundances}

\maketitle

%

\section{Introduction}

In the quest for high-redshift galaxies and their properties in the early 
Universe, the most powerful optical, radio, and space telescopes combined
with efficient instruments are employed to collect the tiny light emission of 
these very faint targets. One of the classes of high-redshift galaxies which 
has been studied in more details is the class of the so-called ``Lyman break 
galaxies''. 

Lyman break galaxies (LBGs) are UV-selected galaxies characterized by a break 
in their ultraviolet continuum, that is due to the Lyman limit from 
intergalactic and interstellar (within the galaxy) \ion{H}{i} absorption below
912~\AA. They are thus easily found using the color-color technique 
\citep{steidel96}. This technique is particularly efficient at $z>2$, where 
the absorption from the intergalactic medium (IGM) is more pronounced and the 
galaxy UV flux is redshifted to optical, allowing observations from the ground. 
Thousands of LBGs have now been discovered, they are the most common galaxy 
population detected at $z\sim 2-3$. While many of the global properties of 
these galaxies, such as their luminosity function, clustering, large-scale 
distribution, and contribution to the star formation rate density of the 
Universe, are in the process of being well understood 
\citep[e.g.,][]{giavalisco98,adelberger03,erb06a,law07,reddy08}, there have 
been few detailed spectroscopic studies to date on their individual properties, 
such as their stellar populations, and chemical enrichment and kinematics of 
their interstellar medium (ISM).

The limited access to individual properties of LBGs is the result of the
faintness of these $L^*$ galaxies: their apparent optical magnitudes fainter
than $R\simeq 24.5$ at $z\sim 3$ make spectroscopic observations challenging. 
The high-resolution spectroscopy, in particular, appears as hardly achievable 
until the 30~m-class telescopes will come into operation. Nevertheless, 
rest-frame UV and optical, low-resolution composite spectra of LBGs have begun 
to provide some insights into the physical properties of these high-redshift 
galaxies. It appears that LBGs resemble present-day star-forming galaxies with 
spectra characterized by young and massive stellar populations of near-solar
metallicities, dominated by on-going star formation, with strong outflows, and
dusty component within abundant neutral gas \citep[e.g.,][]{rix04,shapley03,
shapley04,erb06b}.

For a few objects, nature provides an alternative route to bypass the step of 
next generation telescopes with large collecting areas and study individual 
LBGs spectroscopically at medium and high resolutions. This happens in case of 
fortuitous alignments of LBGs with foreground mass concentrations which lead to 
light magnification due to gravitational lensing. The best-known example is the 
LBG MS\,1512--cB58 (cB58) that is exceptionally bright for its redshift 
$z=2.73$, benefiting of a lensing magnification factor of $\sim 30$ 
\citep[$g=21.08$, $r=20.60$;][]{ellingson96}. This allowed a uniquely detailed 
chemical and kinematical analysis of the ISM, stars, \ion{H}{ii} regions, 
Ly$\alpha$ profile, and surrounding IGM of this high-redshift galaxy 
\citep{pettini00,pettini02,savaglio02,teplitz00,schaerer08,rix04}. 

Recently, new search techniques for strongly-lensed high-redshift galaxies, 
mainly based on the Sloan Digital Sky Survey (SDSS), have yielded additional 
LBG candidates \citep[see e.g.,][and references therein]{kubo09}. All these
objects are excellent targets for follow-up observations with 
intermediate-to-high resolution spectrographs in the optical and/or in the
near-infrared (NIR). So far, \citet{quider09a} provided a first similar 
in-depth study as the one of cB58 of the lensed LBG the ``Cosmic Horseshoe'' at 
$z=2.38$ \citep[J1148+1930;][]{belokurov07}, thanks to rest-frame UV spectra 
obtained at intermediate-resolution with the ESI/Keck\,II spectrograph. And 
similarly, \citet{quider09b} published the detailed analysis of the rest-frame 
UV spectra of the ``Cosmic Eye'', a lensed LBG at $z=3.07$ 
\citep[J213512.73--010143;][]{smail07}. \citet{hainline09}, on the other hand, 
led a very detailed study of the rest-frame optical spectra of three lensed 
LBGs---the Cosmic Horseshoe, the Clone at $z=2.00$ \citep{lin09}, and 
SDSS\,J0901+1814 at $z=2.26$ \citep{diehl09}---obtained at moderate-resolution 
with the NIRSPEC/Keck\,II spectrograph. Finally, \citet{cabanac08} undertook a 
detailed analysis of the rest-frame UV spectra of the lensed LBG 
FOR\,J0332--3557 at a higher redshift $z=3.77$ \citep{cabanac05}, acquired with 
the FORS2/VLT spectrograph, at a resolution about 4 times lower than that of 
cB58, Cosmic Horseshoe, and Cosmic Eye spectra.

\begin{figure}
\centering
\includegraphics[width=9cm,clip]{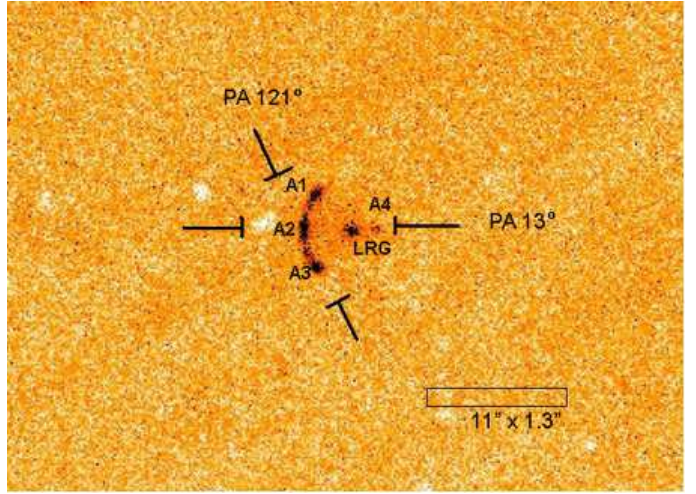}
\caption{45~s X-shooter acquisition image of the 8~o'clock arc obtained through 
the \textit{g'} SDSS filter. The two slit orientations selected for the 
8~o'clock arc observations are shown (see Table~\ref{tab:observations}). The 
various lens images are labeled according to \citet{allam07}. The size of the 
slit we used for the observations with the UV-B spectrograph is also plotted. 
In this good seeing ($0.6''$) image, emission along the whole arc---in between 
the main knots A1, A2, and A3---is clearly visible. Diffuse emission is also 
seen all around the lens galaxy and, in particular, in the region having the 
arc as the outer boundary.}
\label{fig:acquisition}
\end{figure}

The lensed Lyman break galaxy the ``8~o'clock arc'', discovered by 
\citet{allam07} in the SDSS Data Release 4, is another very exciting example of 
high-redshift galaxies which individual physical properties can be studied in 
great details. Indeed, the 8~o'clock arc at $z\simeq 2.73$ is even the 
brightest LBG currently known, with an apparent brightness 3 times higher than 
the one of cB58 ($g=19.95$, $r=19.22$). It is strongly lensed by the $z=0.38$ 
luminous red galaxy (LRG) SDSS\,J002240.91+143110.4, resulting in a total 
magnification factor $\mu = 12.3^{+15}_{-3.6}$. Even after accounting for this 
magnification, the 8~o'clock arc is intrinsically more luminous by about 
2.6~mag (a factor of $\sim 11$) than typical $L^*$ LBGs. The lensing distorts 
the galaxy into four separate knots. Three of them A1, A2, and A3 form a 
partial Einstein ring of radius $\theta_{\rm E} = 3.32''\pm 0.16''$, subtending 
an angle of $162^{\circ}$ and extending over $9.6''$ in length (see 
Fig.~\ref{fig:acquisition}). \citet{finkelstein09} reported the first study of 
low-resolution rest-frame UV and optical spectra of this newly discovered LBG 
obtained with the LRIS/Keck\,I and NIRI/Gemini North spectrographs, 
respectively. They derived several important physical quantities, such as a 
metallicity of $\sim 0.8~Z_{\odot}$ from \ion{H}{ii} regions, a dust extinction 
of $A_{5500} = 1.17\pm 0.36~{\rm mag}$, a stellar mass of $\sim 4.2\times 
10^{11}$~M$_{\odot}$, and a star formation rate of $\sim 270$ 
M$_{\odot}$~yr$^{-1}$.

The 8~o'clock arc was chosen as a target for the first Commissioning period of 
the new X-shooter spectrograph on the Very Large Telescope (VLT). High-quality 
rest-frame UV spectra were acquired of this lensed LBG at 
intermediate-resolution. These spectra are complementary to those acquired by 
\citet{finkelstein09}. Besides MS\,1512--cB58, the Cosmic Horseshoe, and the 
Cosmic Eye, with the 8~o'clock arc we provide the fourth \textit{unprecedented} 
detailed study of stars and ISM gas, from the rest-frame UV spectrum analysis, 
of a high-redshift Lyman break galaxy. The build-up of a sample of such 
comprehensive studies is necessary to better understand the physical properties 
of these high-redshift galaxies and to determine how typical is the well-know 
case of cB58. 

The layout of the paper is as follows. In Sect.~\ref{sect:observations} we 
report on properties of the new X-shooter instrument, on observations, and on 
data reduction procedures. In Sect.~\ref{sect:stellarspectrum} we discuss the 
stellar spectrum of the galaxy, and determine the systemic redshift and the 
metallicity of OB stars. In Sect.~\ref{sect:ISMspectrum} we present the 
interstellar spectrum, and derive the ion column densities. In
Sect~\ref{sect:Lyalpha} we analyze and model the Ly$\alpha$ line profile. 
Finally, in Sect.~\ref{sect:discussion} we summarize the results and discuss 
them in the context of properties of other Lyman break galaxies. Throughout the 
paper, we adopt the standard cosmology with $H_0 = 70$ km~$^{-1}$~Mpc$^{-1}$, 
$\Omega_{\rm M} = 0.3$, and $\Omega_\Lambda = 0.7$.

%

\begin{table*}
\caption{X-shooter observations of the 8~o'clock arc.}             
\label{tab:observations}      
\centering          
\begin{tabular}{c c l c c l}     
\hline\hline       
 & Observing date & Arm & Resolution                 & Exposure time & Slit position\,$^{\rm a}$ \\
 &                &     & $R=\lambda/\Delta \lambda$ & (s)           & \\
\hline
\#\,1 & Nov.\,16, 2009 & UV-B  & 4000 & 3600 & ${\rm PA} = 118^{\circ}$ along A2 and A3 \\
\#\,2 & Nov.\,16, 2009 & VIS-R & 6700 & 3600 & Same as \#\,1 \\
\#\,3 & Nov.\,16, 2009 & UV-B  & 4000 & 3600 & ${\rm PA} = 118^{\circ}$ along A2 and A3 \\
\#\,4 & Nov.\,16, 2009 & VIS-R & 6700 & 3600 & Same as \#\,3 \\
\#\,5 & Nov.\,18, 2009 & UV-B  & 4000 & 4500 & ${\rm PA} = 121^{\circ}$ along A2 and A3 \\
\#\,6 & Nov.\,18, 2009 & VIS-R & 6700 & 4500 & Same as \#\,5 \\
\#\,7 & Nov.\,19, 2009 & UV-B  & 4000 & 4500 & ${\rm PA} = 13^{\circ}$ along A2 and the LRG \\	    
\#\,8 & Nov.\,19, 2009 & VIS-R & 6700 & 4500 & Same as \#\,7
\\
\hline                  
\end{tabular}
\begin{minipage}{125mm}
\smallskip
$^{\rm a}$ Labels of the lens images as in \citet{allam07} and 
Fig.~\ref{fig:acquisition}.
\end{minipage}
\end{table*}
%

%

\section{Observations and data reduction}
\label{sect:observations}

X-shooter is the first of the $2^{\rm nd}$ generation VLT instruments 
\citep{dodorico06} currently mounted on the 8.2~m Kueyen telescope at Cerro 
Paranal, Chile. It was built by a consortium of institutes in Denmark, France, 
Italy, and The Netherlands, and by the European Southern Observatory (ESO) 
which was responsible for its final integration and installation at the 
telescope. It consists of three echelle spectrographs with prism 
cross-dispersion mounted on a common structure at the Cassegrain focus. The 
light beam from the telescope is split in the instrument by two dichroics which 
direct the light in the spectral ranges of $300-560$ nm and $560-1015$ nm to 
the slit of the ultraviolet-blue (UV-B) and visual-red (VIS-R) spectrographs, 
respectively. The undeviated beam in the spectral range of $1025-2400$ nm feeds 
the near-infrared (NIR) spectrograph. The UV-B and VIS-R spectrographs operate 
at ambient temperature and pressure and deliver 2D spectra on $4{\rm k}\times 
2{\rm k}$ pixels CCDs. The NIR spectrograph is enclosed in a vacuum vessel and 
kept at a temperature of $\sim 80$~K by a continuous flow of liquid nitrogen. 
The spectral format in the three spectrographs is fixed, with the possibility 
to change the resolution by using slits of different widths. 

After two Commissioning runs with the UV-B and VIS-R spectrographs in November 
2008 and January 2009, the instrument is operating in its full configuration, 
including the NIR spectrograph, from March 2009. Calibrations and data of 
scientific value obtained during the Commissioning runs are part of the public
release by the ESO 
archive\footnote{http://www.eso.org/sci/facilities/develop/instruments/xshooter/X-shooter\_CommDataRelease\_text.html}. 
The data used in this paper were obtained during the first Commissioning run of 
November 2008. These were the first nights of the instrument at the telescope, 
when the Commissioning team was testing the observing procedures and the 
instrument behavior on sky targets. The 8~o'clock arc was considered as a good 
test-case of faint galaxy observations. While the instrument set-up, the 
observing strategy, and the exposure times were not optimized yet, the acquired 
spectra are still mostly of good quality and provide unique data of the 
8~o'clock Lyman Break galaxy. 

The list of observations is summarized in Table~\ref{tab:observations}. A total
exposure time of 16\,200~s (2 exposures of 3600~s and 2 others of 4500~s) was 
obtained on the 8~o'clock arc in good conditions, with a clear sky and seeing 
between $0.6''$ and $1.2''$. Slit widths of $1.3''$ in the UV-B and $1.2''$ in 
the VIS-R were used, corresponding to resolutions $R=4000$ and $R=6700$, 
respectively. Two main slit orientations were selected aligned at position 
angles ${\rm PA} \simeq 121^{\circ}$ and $13^{\circ}$, along the 8~o'clock 
knots A2 and A3 and along the knot A2 and the lens galaxy, respectively. 
Figure~\ref{fig:acquisition} shows the two slit positions on the sky on a 45~s 
acquisition image obtained with X-shooter through the \textit{g'} SDSS filter. 
All the observations were conducted at air mass of $\sim 1.3$.

The UV-B exposures \#\,1, \#\,3 and \#\,5 show an in-focus light ghost which at 
wavelengths below 380 nm partly overlaps the signal of the 8~o'clock knots A2 
and A3, and below 460 nm prevents an accurate sky subtraction in exposures \#\,3 
and \#\,5. After excluding possible contamination from some parasitic light
sources in the instrument or the telescope, the bright lensing LRG, at less 
than 3 arcsec off the slit, was suspected as the source of the ghost. This is
supported by the exposure \#\,7, taken with the slit aligned along the knot A2 
and the LRG, which does not show any light ghost. However, the effect was not 
observed with any other target, and various tests made with bright stars close 
to the slit failed to reproduce it. The matter is still under investigation by 
the instrument team. 

The X-shooter spectra were reduced with the Beta version of the X-shooter 
reduction pipeline \citep{goldoni06} running at ESO. Pixels in the 2D echelle 
format frames are mapped in the wavelength space using calibration frames. Sky 
emission lines are normally subtracted before any resampling, using the method 
developed by \citet{kelson03}. The different orders are then extracted, 
rectified, wavelength calibrated with a constant spectral bin, and merged with 
a weighted average used in the overlapping regions. 

Given the complexity of the geometry of the 8~o'clock arc target and the 
related observations where one single slit is aligned along two knots, we had 
to perform the object extraction and sky subtraction manually. For this 
purpose, we used the 2D wavelength calibrated, resampled, background 
subtracted, order merged UV-B and VIS-R spectra as produced by the pipeline, 
and made manual object and sky extractions by carefully selecting the 
respective extraction windows. The separate extraction of the knots A2 and A3 
in exposures \#\,1 to \#\,6 led to very low signal-to-noise ratio (S/N) 
individual spectra. Since no difference is observed between the A2 and A3 
spectra within the limits of the noise and the signal of the two knots in the 
slit is partly blended, we handled the sum of the signals of the two knots. In 
the case of exposures \#\,7 and \#\,8, where the slit was aligned along the 
knot A2 and the LRG, the extracted object spectrum corresponds solely to the 
knot A2. Exposures with this specific slit orientation were extremely useful to 
estimate possible light contamination of the 8~o'clock arc by the lensing 
galaxy. Indeed, they show a residual emission relative to the sky signal in 
between the LRG and the knot A2. The extracted light profile along the $11''$ 
slit from the nucleus of the LRG up to the sky beyond the knot A2 (see 
Fig.~\ref{fig:acquisition}) follows the de Vaucouleurs law with the typical 
$R^{1/4}$ intensity profile of elliptical galaxies (where $R$ is the distance
relative to the LRG). A significant wavelength dependent contamination by 
the LRG (increasing toward longer wavelengths), of the order of $\sim 30$\%, is 
observed at the position of the 8~o'clock arc. We thus subtracted this light 
contamination in addition to the manually extracted sky signal (as determined 
at larger distances from the lens) from the extracted signals of the knots A2 
and A3 in all exposures. The 1D science spectra from the different exposures 
were then co-added using their S/N as weights\footnote{Solely exposures \#\,1 
and \#\,7 were co-added in the UV-B, because sky subtraction failed in 
exposures \#\,3 and \#\,5 due to the presence of the light ghost (see above).}. 
For this, the wavelength scale was first converted to the vacuum-heliocentric 
scale. The spectra were finally normalized by smoothly connecting regions free 
from absorption features with a spline function. A particularly careful 
continuum fitting was done in the Ly$\alpha$ forest. 

The final UV-B and VIS-R spectra have, respectively, a S/N per resolution 
element of $\sim 10$ from 440 to 560 nm and $\sim 13$ from 560 to 800 nm. From 
the widths of the sky emission lines, we measured a spectral resolution of 1.25 
\AA\ (69 km~s$^{-1}$) FWHM at 540 nm in the UV-B spectrum and 0.85 \AA\ (42
km~s$^{-1}$) FWHM at 600 nm in the VIS-R spectrum. This is in very good 
agreement with the resolutions expected with the slit widths used for our 
observations. The sky emission lines allowed us also to check the wavelength 
calibration of our X-shooter spectra. The achieved wavelength calibration
accuracy is $\sim 10$ km~s$^{-1}$ rms in both the UV-B and VIS-R spectra.

%

\section{The stellar spectrum}
\label{sect:stellarspectrum}

The rest-frame UV spectrum of the 8~o'clock arc consists of the integrated 
light from the hot and luminous O and B stars in the galaxy (the
\textit{stellar} spectrum) on which are superposed the resonant absorption 
lines produced by the interstellar gas (the \textit{interstellar} spectrum). 
The careful analysis of all these resulting lines brings precious information 
on the physical properties of both the stars and gas in this high redshift 
galaxy. On top of these lines, absorption from the Ly$\alpha$ forest and 
several intervening metal-line systems is also found. We discuss first the 
stellar spectrum.

%

\begin{table*}
\caption{Stellar photospheric absorption lines and emission lines.}             
\label{tab:stellarlines}      
\centering          
\begin{tabular}{l c l c c c  D{.}{.}{3}  D{.}{.}{3}}     
\hline\hline       
Ion & $\lambda_{\rm lab}$$^{\rm a}$ & Origin & $z_{\rm stars}$$^{\rm b}$ & $z_{\rm em}$$^{\rm c}$ & \multicolumn{1}{c}{$\Delta\upsilon$$\,^{\rm d}$} & \multicolumn{1}{c}{$W_0$$^{\rm e}$} & \multicolumn{1}{c}{$\sigma (W_0)$$\,^{\rm e}$} \\
    & (\AA)           	   	    &        &                           &                        & \multicolumn{1}{c}{(km~s$^{-1}$)}                & \multicolumn{1}{c}{(\AA)}           & \multicolumn{1}{c}{(\AA)} \\
\hline
\ion{Si}{iii}    & 1294.543 & stars		  & 2.7349 &        & $-165$ to $+105$ &  0.13           & 0.02 \\
\ion{C}{iii}     & 1296.330 & stars		  & 2.7349 &        & $-165$ to $+195$ &  0.16\,^{\rm f} & 0.02\,^{\rm f} \\
\ion{Si}{iii}    & 1296.726 & stars		  & 2.7348 &        & $-165$ to $+195$ &  0.16\,^{\rm f} & 0.02\,^{\rm f} \\
\ion{C}{ii}      & 1323.929 & stars		  & 2.7348 &        & $-165$ to $+195$ &  0.21\,^{\rm g} & 0.03\,^{\rm g} \\
\ion{N}{iii}     & 1324.316 & stars		  & 2.7345 &        & $-165$ to $+195$ &  0.21\,^{\rm g} & 0.03\,^{\rm g} \\
\ion{Si}{ii}$^*$ & 1533.431 & recombination	  &        & 2.7352 & $~-95$ to $+185$ & -0.20           & 0.02 \\   
\ion{N}{iv}      & 1718.551 & stars		  & 2.7347 &        & $-210$ to $+165$ &  0.57           & 0.06 \\
$[$\ion{C}{iii}] & 1906.683 & \ion{H}{ii} regions &        & 2.7348 & $-185$ to $+165$ & -0.35           & 0.04 \\   
\ion{C}{iii}]    & 1908.734 & \ion{H}{ii} regions &        & 2.7352 & $-165$ to $+185$ & -0.26           & 0.03	  
\\
\hline                  
\end{tabular}
\begin{minipage}{118mm}
\smallskip
$^{\rm a}$ Vacuum wavelengths. \\
$^{\rm b}$ Vacuum heliocentric and values measured from the centroid wavelength 
of the photospheric absorption lines. \\
$^{\rm c}$ Vacuum heliocentric and values measured from the Gaussian fits to 
the emission lines. \\
$^{\rm d}$ Velocity range for equivalent width measurements relative to 
$z_{\rm sys} = 2.7350$. \\
$^{\rm e}$ Rest-frame equivalent width and $1\,\sigma$ error. \\
$^{\rm f}$ This value refers to the blend of lines 
\ion{C}{iii}\,$\lambda$1296.330 and \ion{Si}{iii}\,$\lambda$1296.726. \\
$^{\rm g}$ This value refers to the blend of lines 
\ion{C}{ii}\,$\lambda$1323.929 and \ion{N}{iii}\,$\lambda$1324.316.
\end{minipage}
\end{table*}
%

\begin{figure}
\centering
\includegraphics[width=7cm,clip]{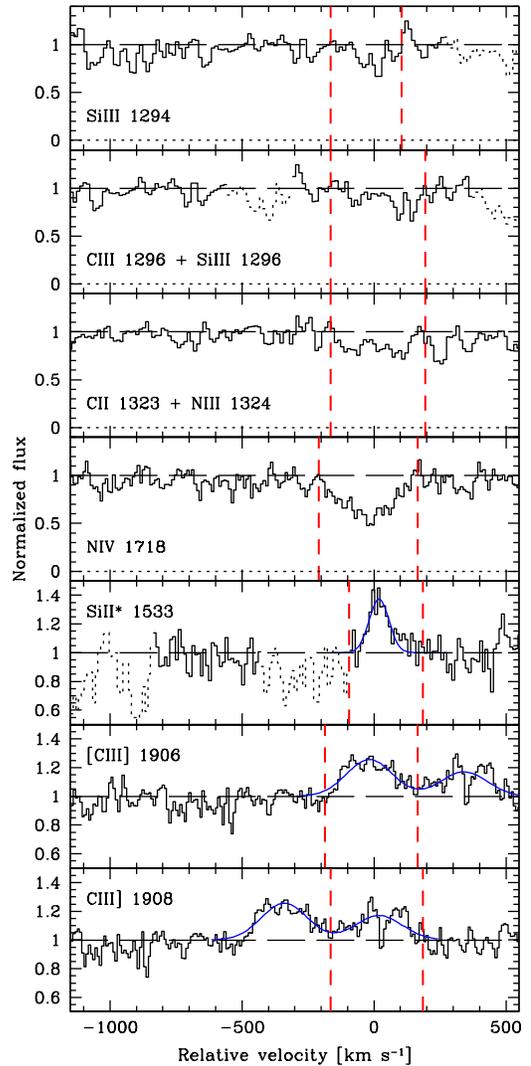}
\caption{Velocity plots of normalized profiles of the stellar photospheric 
absorption lines and the emission lines detected in the 8~o'clock arc. The zero 
velocity is fixed to $z_{\rm sys} = 2.7350$. The vertical red dashed lines 
indicate the velocity range over which the equivalent widths listed in 
Table~\ref{tab:stellarlines} are measured. The blue solid lines are Gaussian
fits to the emission lines. The dotted lines are used to indicate blends with
absorption features that are \textit{not} those labeled in each panel.}
\label{fig:stellarlines}
\end{figure}
%

\subsection{Systemic redshift}
\label{sect:zsys}

Most of the low-contrast/weak structure seen in the continuum of high S/N 
spectra of star-forming galaxies is caused by stellar features, not noise. 
These features are largely blends of different stellar lines which require 
stellar population synthesis to be analyzed quantitatively (see 
Sect.~\ref{sect:stellarlines}). \citet{pettini00,pettini02} identified a few 
stellar photospheric lines that appear to be least affected by blending, and 
can therefore provide a measure of the \textit{systemic} redshift of the 
stellar population.

We identified 6 of these photospheric absorption lines. They are listed in 
Table~\ref{tab:stellarlines} with their redshifts and rest-frame equivalent 
widths, $W_0$, and plotted in velocity space in Fig.~\ref{fig:stellarlines}. 
The tabulated redshifts\footnote{All redshifts quoted in this paper are vacuum 
heliocentric.} are those derived from the centroid of the lines, that 
corresponds to the mean wavelength of the line weighted by the absorption at 
each wavelength. 
Solely the redshift derived from the \ion{N}{iv}\,$\lambda$1718 line provides a 
reliable measure of $z_{\rm stars} = 2.7347\pm 0.0003$, given the larger 
strength of this line with respect to the other stellar photospheric absorption 
lines only marginally detected ($2-4\,\sigma$) and the blends of \ion{C}{iii} 
with \ion{Si}{iii} and \ion{C}{ii} with \ion{N}{iii}. 

Also included in Table~\ref{tab:stellarlines} are 3 emission lines: the nebular 
\ion{C}{iii}]\,$\lambda\lambda$1906,\,1908 doublet clearly resolved in our 
X-shooter spectrum of the 8~o'clock arc and the fine-structure 
\ion{Si}{ii}$^*$\,$\lambda$1533 emission line (Fig.~\ref{fig:stellarlines}). 
The mean $z_{\rm em} = 2.7351\pm 0.0002$ is in very good agreement with 
$z_{\rm stars}$, differing by only 32 km~s$^{-1}$. We thus adopt throughout the 
paper $z_{\rm sys} = 2.7350\pm 0.0003$---the mean of $z_{\rm stars}$ and 
$z_{\rm em}$---as the systemic redshift in the 8~o'clock arc. 

\citet{finkelstein09} measured $z_{\rm H\,\mathsc{ii}} = 2.7333\pm 0.0001$ from 
the mean of 6 well-detected emission lines formed in \ion{H}{ii} regions 
redshifted into the NIR. The large difference, $\Delta \upsilon = -136$ 
km~s$^{-1}$, between $z_{\rm H\,\mathsc{ii}}$ and $z_{\rm sys}$ derived here 
(based also on two nebular \ion{C}{iii}] emission lines) is difficult to 
explain. Moreover, their redshift estimates from the stellar photospheric and
interstellar lines redshifted into the optical show the same discrepancy. The
higher resolution and wavelength accuracy of our X-shooter spectra than of 
their observations lead us to preferentially trust our redshift estimates.

%

\subsection{Photospheric absorption lines}
\label{sect:stellarlines}

Similarly to \citet{pettini00}, \citet{cabanac08}, and more recently to 
\citet{quider09a,quider09b} who performed the stellar modeling of the 
integrated spectra of four lensed LBGs, respectively, we analyze the 
photospheric and wind lines in the spectrum of the 8~o'clock arc. For this, we 
compare our data with the synthetic spectra computed by \citet{rix04} using 
their updated version of the population synthesis code \textit{Starburst99} 
\citep{leitherer99,leitherer01} which couples libraries of theoretical UV OB 
stellar spectra with stellar evolutionary tracks. We follow their assumption of 
a continuous star formation mode---with a Salpeter initial mass function (IMF) 
between 1 and 100 M$_{\odot}$ and a constant star formation rate for 100 
Myr---which seems to be the better description for most LBGs. 

\citet{leitherer01} and \citet{rix04} explored in detail a number of spectral 
regions in the integrated UV stellar spectra of star-forming galaxies that are 
sensitive to metallicity and clean of other spectral features. They identified 
three very promising metallicity indicators: (1)~the ``1370'' index which 
arises from the blending of the \ion{O}{v}\,$\lambda$1371 and 
\ion{Fe}{v}\,$\lambda\lambda$1360,1380 absorption lines over 1360--1380 \AA; 
(2)~the ``1425'' index which arises from the blending of the 
\ion{Si}{iii}\,$\lambda$1417, \ion{C}{iii}\,$\lambda$1427 and 
\ion{Fe}{v}\,$\lambda$1430 absorption lines over 1415--1435 \AA; and (3)~the 
``1978'' index which arises from the blending of numerous \ion{Fe}{iii}
transitions over 1935--2020 \AA. The equivalent widths of all these indices 
increase monotonically with metallicity. The ``1978'' index is particularly 
interesting, because it has a larger equivalent width and is free of 
contaminating interstellar lines.

To compare our high-resolution X-shooter spectra of the 8~o'clock arc with the 
synthetic stellar spectra of \citet{rix04}, we smoothed our non-normalized 
spectra to the resolution ${\rm FWHM} = 2.5$ \AA\ of the synthetic spectra (in 
the rest-frame wavelength scale) by convolution with Gaussian profiles of the 
appropriate widths. We then normalized the smoothed X-shooter spectra by 
division by a spline curve through the mean flux in each of the 
pseudo-continuum windows deemed to be free of absorption/emission features 
identified by \citet[][Table~3]{rix04}. Given the broad and shallow nature of 
the photospheric blends making up the ``1370'', ``1425'', and ``1978'' indices, 
their equivalent widths are very sensitive to the continuum normalization, 
which in turn depends on the spectral resolution of the spectra. It is 
therefore mandatory to apply the above steps on the acquired data.

\begin{figure*}
\centering
\includegraphics[width=15cm,clip]{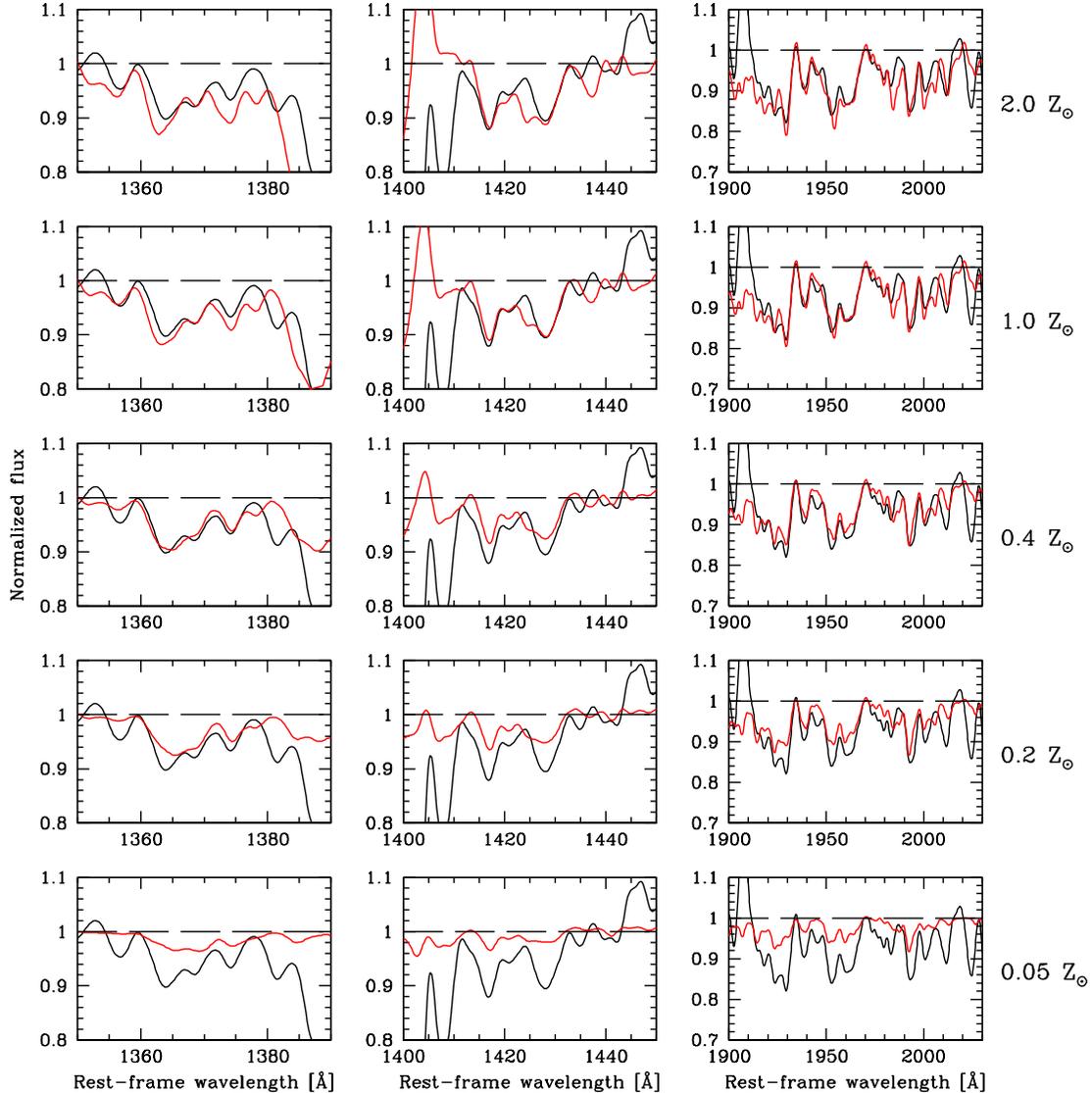}
\caption{Comparison of the X-shooter spectra of the 8~o'clock arc smoothed to 
the ${\rm FWHM} = 2.5$ \AA\ resolution (black) with synthetic stellar spectra 
(red) produced by \citet{rix04} for 5 metallicities, from twice solar to 1/20 
of solar, as indicated. Wavelength portions encompassing the ``1370'', 
``1425'', and ``1978'' metallicity indices are plotted from the left to the
right, respectively. The strong absorption between 1400 and 1410 \AA\ 
corresponds to the interstellar \ion{Si}{iv}\,$\lambda$1402 line plus an 
intervening metal-line and the emission at $\sim 1907$ \AA\ to the blend of the 
nebular \ion{C}{iii}]\,$\lambda\lambda$1906,\,1908 doublet, that are 
unaccounted for in the Rix et~al.\ models.} 
\label{fig:stellarmodels}
\end{figure*}

In the panels of Fig.~\ref{fig:stellarmodels} are shown portions of the 
smoothed X-shooter spectra of the 8~o'clock arc with the synthetic stellar 
spectra of \citet{rix04} for 5 values of metallicity, from 1/20 of solar to 
twice solar, in the 1350--1390 \AA\ (left), 1400--1450 \AA\ (middle), and 
1900--2030 \AA\ (right) wavelength intervals, corresponding to the ``1370'', 
``1425'', and ``1978'' metallicity indices, respectively. The synthetic spectra 
do a remarkably good job at reproducing the observed spectra; except for the 
strong absorption between 1400 and 1410 \AA\ which corresponds to the 
interstellar \ion{Si}{iv}\,$\lambda$1402 line plus an intervening metal-line 
and the emission at $\sim 1907$ \AA\ which is the blend of the nebular 
\ion{C}{iii}]\,$\lambda\lambda$1906,\,1908 doublet, that are unaccounted for in 
the Rix et~al.\ models. Among the 5 metallicities considered, $Z=1~Z_{\odot}$ 
and $0.4~Z_{\odot}$ are those that most closely match the observations in all 
the ``1370'', ``1425'', and ``1978'' regions. Using the relations of 
\citet{rix04} between the equivalent widths of the ``1425'' and ``1978'' line 
blends and $\log(Z/Z_{\odot})$, we get a first estimate of the metallicity of 
the 8~o'clock LBG: the measured $W_0(1425) = 1.16$ \AA\ over the 1415--1435 
\AA\ interval gives $Z=0.85~Z_{\odot}$, and the measured $W_0(1978) = 5.56$ 
\AA\ over the 1935--2020 \AA\ interval gives $Z=0.79~Z_{\odot}$. The two 
metallicities are in excellent agreement.

The P-Cygni lines formed in the expanding winds of the most luminous OB stars
are other very important metallicity indicators. The optical depth of these 
lines is sensitive to the mass-loss rate which in turn decreases with 
decreasing metallicity. The \ion{C}{iv}\,$\lambda\lambda$1548,\,1550 doublet is 
the strongest P-Cygni line covered by the X-shooter spectra of the 8~o'clock 
arc. We do not consider the \ion{C}{iv} doublet here as a metallicity 
indicator, as no precise metallicity calibration exists for these lines so far. 
The stellar models of \citet{rix04} are not applicable, because at the 
resolution ${\rm FWHM} = 2.5$ \AA\ of their synthetic spectra the 
interpretation of the \ion{C}{iv} P-Cygni line is complicated by its blending 
with the interstellar \ion{C}{iv} doublet and to some extent the 
\ion{Si}{ii}\,$\lambda$1526 absorption. 

%

%

\subsection{Emission lines}
\label{sect:emissionlines}

As did \citet{quider09a} in the Cosmic Horseshoe, we detect the nebular 
\ion{C}{iii}]\,$\lambda\lambda$1906,\,1908 doublet formed in \ion{H}{ii} 
regions that is clearly resolved in our X-shooter spectra of the 8~o'clock arc 
(see Fig.~\ref{fig:stellarlines}). The mean redshift, $z_{\rm C\,\mathsc{iii}]} 
= 2.7350$, is in very good agreement with both the redshift of the stellar 
photospheric absorption lines, $z_{\rm stars}$, and the redshift of another 
detected emission line, \ion{Si}{ii}$^*$\,$\lambda$1533 (see 
Table~\ref{tab:stellarlines}). According to \citet{pettini00}, \ion{Si}{ii}$^*$ 
can be interpreted as a recombination line to the fine-structure level of the 
ground state of \ion{Si}{ii}, presumably arising in an \ion{H}{ii} region. We
also searched for nebular \ion{O}{iii}]\,$\lambda\lambda$1660.809,1666.150
emission, but these lines appear to be below the detection limit of our data. 

Applying Gaussian fits to the \ion{C}{iii}] doublet allows to derive the full
width at half maximum, FWHM, and hence the velocity dispersion, $\sigma = 
{\rm FWHM}/2.355\times c/\lambda_{\rm obs}$, which is a measure of the dynamics 
of the gas bound to the galaxy by gravity. We obtain very consistent velocity 
dispersions for the two \ion{C}{iii}] lines, $\sigma_{\rm C\,\mathsc{iii}]} = 
70$ and 71 km~s$^{-1}$, respectively (after correcting for the instrumental 
resolution). The mean value is in very good agreement with the velocity 
dispersions obtained by \citet{hainline09} for the few lensed LBGs known so 
far, while it is about twice the velocity dispersion obtained by 
\citet{finkelstein09} for the nebular H$\alpha$ line redshifted into the NIR. 

The ratio of the two \ion{C}{iii}] lines is a function of the electron density. 
It varies from values of about 1.5 to 0.8 in the range of $n(e) = 100$ to
30\,000 cm$^{-3}$, respectively. The measured 
\ion{C}{iii}]\,$\lambda$1906/$\lambda$1908 line ratio of $\sim 1.5$ in the 
8~o'clock arc points to the lowest end of electron densities, $n(e) \sim 100$. 
This is consistent with the values usually observed in local star-forming 
galaxies, but appears lower than the electron densities determined from the 
nebular [\ion{S}{ii}]\,$\lambda\lambda$6717,\,6731 doublet in the other 
high-redshift LBGs \citep{brinchmann08,hainline09}.

\begin{figure}
\centering
\includegraphics[width=9cm,clip]{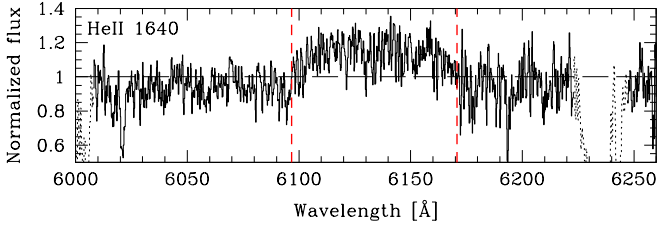}
\caption{Normalized profile of the \ion{He}{ii}\,$\lambda$1640 emission line
detected in the 8~o'clock arc. The vertical red dashed lines indicate the
wavelength range over which the equivalent width is measured. The dotted lines
are used to indicate ISM absorption features.} 
\label{fig:HeII1640}
\end{figure}

Interestingly, we also detect another very broad emission feature in the
8~o'clock arc, extending over $\sim 75$ \AA\ (${\rm FWHM} \sim 2000$
km~s$^{-1}$), which at the systemic redshift $z_{\rm sys} = 2.7350$ is 
identified as the \ion{He}{ii}\,$\lambda$1640.418 line (see 
Fig.~\ref{fig:HeII1640}). The \ion{He}{ii} emission is known to be a possible 
signature of very massive stars, produced by Wolf-Rayet (WR) stars. Hints for 
this emission were already detected in the composite spectrum of LBGs of 
\citet{shapley03}, and more recently \citet{cabanac08} reported a first clear 
detection in the lensed FOR\,J0332--3557 LBG. 

We measure a rest-frame equivalent width $W_0({\rm He\,\mathsc{ii}}) = -2.45\pm 
0.22$ \AA, very similar to the value obtained by \citet{cabanac08} in 
FOR\,J0332--3557, but somewhat larger than the value of 
$W_0({\rm He\,\mathsc{ii}}) = -1.3\pm 0.3$ \AA\ measured by 
\citet{brinchmann08} in the composite spectrum of $z\sim 3$ LBGs of 
\citet{shapley03}. Comparison with local measurements suggests that the 
equivalent width of \ion{He}{ii}\,$\lambda$1640 in the 8~o'clock arc is similar 
to values found in nearby starburst super-star clusters 
\citep[][Table~3]{chandar04}. Comparison with evolutionary synthesis models 
\citep{schaerer98,brinchmann08} shows that the strength of the observed 
\ion{He}{ii} emission can be understood by either relatively young bursts of 
$\la 10$ Myr, by continuous star formation at solar metallicity over $\la 20$ 
Myr, or by longer star formation timescales at metallicities above solar. We 
conclude that the strong \ion{He}{ii}\,$\lambda$1640 emission line in the 
8~o'clock arc very likely results from WR stars. However, given the 
uncertainties involved in the models (e.g., line luminosity calibrations, 
evolution of massive stars) and in the precise star formation history, we 
cannot draw more solid conclusions from this spectral feature.

%

\section{The interstellar spectrum}
\label{sect:ISMspectrum}

\subsection{Interstellar absorption lines}
\label{sect:ISMlines}

The rest-frame UV spectrum of the 8~o'clock LBG is dominated by interstellar
absorption lines. Table~\ref{tab:ISMlines} lists all the ISM lines detected and 
measured in our X-shooter spectra. Together with Ly$\alpha$ (which will be 
discussed separately in Sect.~\ref{sect:Lyalpha}), we cover 31 transitions of 
elements from H to Zn in a variety of ionization stages from neutral 
(\ion{H}{i}, \ion{O}{i}) to highly ionized species (\ion{C}{iv}, \ion{N}{v}).
The low-ionization features are associated with neutral gas, while the
high-ionization features predominantly trace gas at $T \geq 10^4$ K which is
ionized by a combination of radiation from massive stars and collisional
processes. All the interstellar lines are seen against the continuum provided 
by the integrated light of O and B stars in the galaxy. Vacuum rest-frame 
wavelengths, $\lambda_{\rm lab}$, and oscillator strengths, $f$, of the 
transitions are from the compilation by \citet{morton03} and \citet{jenkins06}. 
The rest-frame equivalent widths, $W_0$, with their $1\,\sigma$ errors were 
measured by summing the absorption over fixed velocity ranges, 
$\Delta\upsilon$. These velocity ranges were chosen to encompass the full 
extent of the absorption while minimizing the amount of continuum included. 

%

%

\begin{figure*}
\centering
\includegraphics[width=15cm,clip]{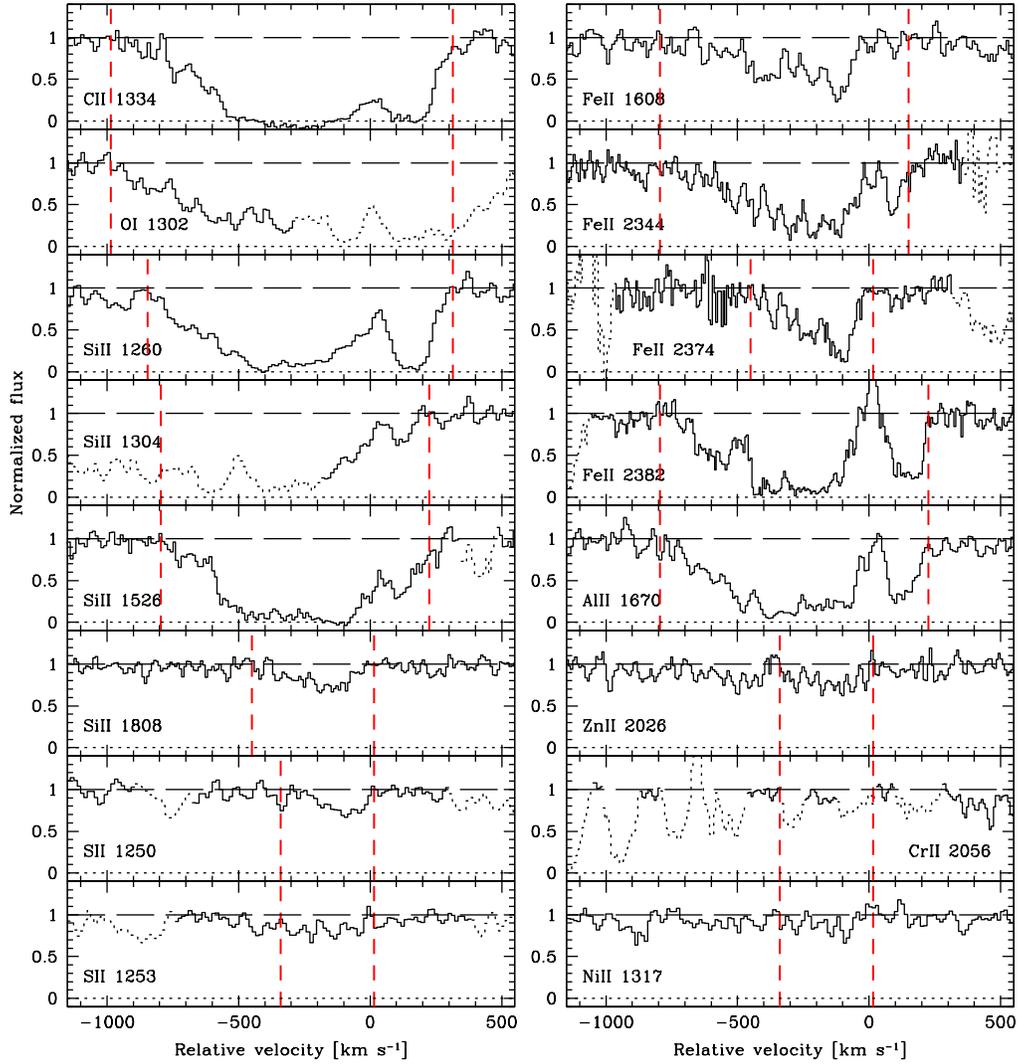}
\caption{Velocity plots of normalized profiles of the low-ionization
interstellar absorption lines detected in the 8~o'clock arc. The zero velocity
is fixed to $z_{\rm sys} = 2.7350$. The vertical red dashed lines indicate the 
velocity range over which the equivalent widths listed in
Table~\ref{tab:ISMlines} are measured. The dotted lines are used to indicate 
blends with absorption features that are \textit{not} those labeled in each 
panel.} 
\label{fig:lowions}
\end{figure*}

\begin{figure*}
\centering
\includegraphics[width=15cm,clip]{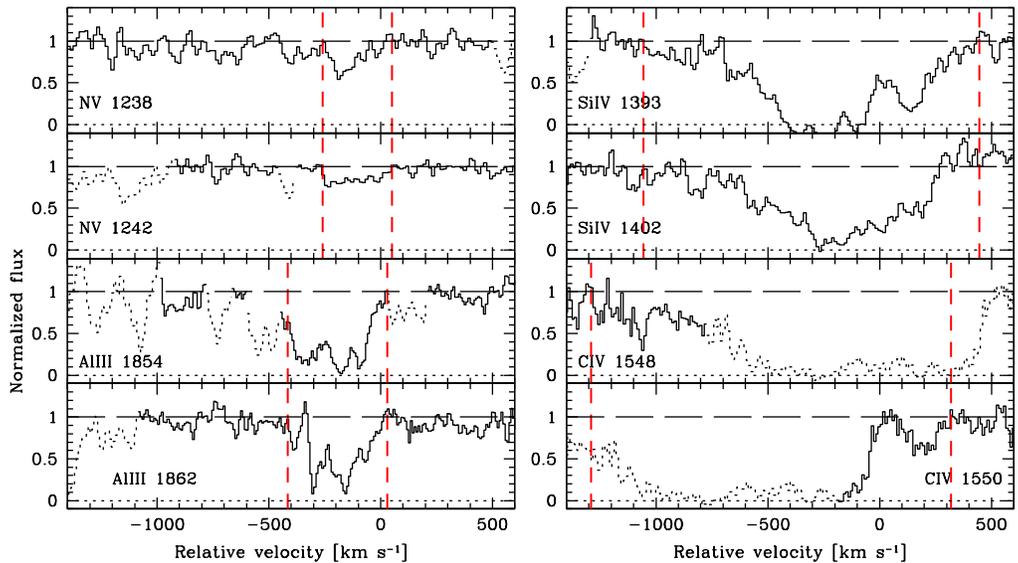}
\caption{Same as Fig.~\ref{fig:lowions}, but for the normalized profiles of the
high-ionization interstellar absorption lines detected in the 8~o'clock arc.}
\label{fig:highions}
\end{figure*}

Figures~\ref{fig:lowions} and \ref{fig:highions} show, respectively, the 
velocity profiles of most of the low- and high-ionization ISM absorption lines
detected. We can note that the ISM lines in the 8~o'clock arc are very broad 
with absorption in the strongest transitions extending over a velocity range 
$\Delta\upsilon \sim 1300$ km~s$^{-1}$, from about $-985$ to +315 km~s$^{-1}$ 
relative to $z_{\rm sys} = 2.7350$. All the ion stages observed show similar 
absorption profiles: (i)~they span the same overall velocity range (except the 
\ion{C}{iv} doublet which is much broader, but it is also much stronger than 
the strongest low-ionization lines), and (ii)~they are characterized by the 
same two main absorption components with the major optical depth extending from 
$-450$ to +50 km~s$^{-1}$ and the minor one being located at about +120 
km~s$^{-1}$. 

Gas at the peak optical depth in the most clearly defined line profiles of the
8~o'clock arc (the unsaturated \ion{Si}{ii}\,$\lambda$1808, 
\ion{Fe}{ii}\,$\lambda$1608, and \ion{Fe}{ii}\,$\lambda$2374 lines) occurs at 
$z_{\rm ISM} = 2.7335\pm 0.0005$. This leads to a \textit{blueshift} of the 
interstellar lines of $\upsilon_{\rm ISM} \simeq -120$ km~s$^{-1}$ with respect 
to the stars and $z_{\rm sys}$. When redshifts of the centroid of the ISM lines 
($z_{\rm abs}$ values listed in Table~\ref{tab:ISMlines}) are considered, they 
give a mean $\langle z_{\rm abs}\rangle = 2.7324\pm 0.0006$ and a velocity 
offset $\upsilon_{\rm ISM} \simeq -209$ km~s$^{-1}$. \citet{finkelstein09} 
derived a similar velocity offset ($-160$ km~s$^{-1}$) relative to the stars 
for the 8~o'clock arc. Such a blueshift of the interstellar gas is a common 
feature of star-forming galaxies at low as well as high redshifts 
\citep{heckman00,pettini01,shapley03,quider09a}. It is generally accepted that 
it results from large-scale outflows of the interstellar medium driven by the 
kinetic energy deposited by supernovae and the winds of massive stars. The 
outflow speed of $-120$ km~s$^{-1}$ is typical of $z\sim 3$ LBGs for which 
\citet{shapley03} derived a mean value of $-150\pm 60$ km~s$^{-1}$. The 
location and nature of the interstellar gas moving at positive velocities 
relative to the stars remain, on the other hand, unexplained.

%

\begin{sidewaystable*}
\caption{Interstellar absorption lines.}
\label{tab:ISMlines}
\centering
\begin{tabular}{l c c c c c c  D{.}{.}{3.3}  c  D{.}{.}{3.3}  c  D{.}{.}{3.3}  c} 
\hline\hline             
Ion & $\lambda_{\rm lab}$$^{\rm a}$ & $f$ & $z_{\rm abs}$$^{\rm b}$ & \phantom{eee}$\Delta\upsilon$$\,^{\rm c}$ & $W_0$$^{\rm d}$ & $\sigma (W_0)\,^{\rm d}$ & 
\multicolumn{1}{c}{$\log N_{\rm aod}$} & \multicolumn{1}{c}{$\sigma (\log N_{\rm aod})$} & \multicolumn{1}{c}{$\log N_{\rm adopt}$$^{\rm e}$} & \multicolumn{1}{c}{$\sigma (\log N_{\rm adopt})$$\,^{\rm e}$} & 
\multicolumn{1}{c}{[X/H]$\,^{\rm f}$} & \multicolumn{1}{c}{$\sigma$([X/H])$\,^{\rm f}$} \\
& (\AA) & & & \phantom{eee}(km~s$^{-1}$) & (\AA) &(\AA) & \multicolumn{1}{c}{(cm$^{-2}$)} & \multicolumn{1}{c}{(cm$^{-2}$)} & \multicolumn{1}{c}{(cm$^{-2}$)} & \multicolumn{1}{c}{(cm$^{-2}$)} & & \\
\hline
\ion{H}{i}       & 1215.6700 & 0.41640 & 	&                   &	   &      &                 &      &  20.57 & 0.10 &        & \\
\ion{O}{i}       & 1302.1685 & 0.04887 & B      & $~-985$ to $+315$ & 3.62 & 0.41 & >15.99\,^{\dag} &      & >15.99 &	   & >-1.24 & \\
\ion{N}{v}       & 1238.8210 & 0.15700 & 2.7334 & $~-260$ to $+50~$ & 0.24 & 0.03 &  14.13          & 0.09 &  14.23 & 0.10 &        & \\
	         & 1242.8040 & 0.07823 & 2.7335 & $~-260$ to $+50~$ & 0.21 & 0.02 &  14.32          & 0.10 &	    &	   &        & \\
\ion{C}{ii}      & 1334.5323 & 0.12780 & 2.7321 & $~-985$ to $+315$ & 4.04 & 1.73 & >15.84\,^{\dag} &	   & >15.84 &	   & >-1.12 & \\
\ion{C}{iv}      & 1548.1949 & 0.19080 & B      & $-1290$ to $+320$ & 5.65 & 2.39 & >15.60\,^{\dag} &	   & >15.93 &	   &        & \\
	         & 1550.7700 & 0.09522 & B      & $-1290$ to $+320$ & 5.84 & 2.46 & >15.93\,^{\dag} &	   &	    &      &        & \\
\ion{Si}{ii}     & 1260.4221 & 1.00700 & 2.7319 & $~-845$ to $+315$ & 3.23 & 0.56 & >14.72\,^{\dag} &	   &  15.89 & 0.10 &  -0.19 & 0.14 \\
	         & 1304.3702 & 0.09400 & B      & $~-795$ to $+225$ & 2.73 & 0.32 & >15.60\,^{\dag} &	   &	    &      &        & \\
	         & 1526.7065 & 0.12700 & 2.7319 & $~-795$ to $+225$ & 3.48 & 1.81 & >15.55\,^{\dag} &	   &	    &      &        & \\
	         & 1808.0129 & 0.00218 & 2.7324 & $~-450$ to $+15~$ & 0.43 & 0.05 &  15.89          & 0.10 &	    &	   &        & \\
\ion{Si}{ii}$^*$ & 1264.7377 & 0.90340 & 2.7326 & $~-450$ to $+15~$ & 0.48 & 0.06 &  13.67          & 0.10 &  13.62 & 0.10 &        & \\
	         & 1309.2758 & 0.14680 & 2.7329 & $~-340$ to $+15~$ & 0.07 & 0.01 &  13.56          & 0.10 &	    &	   &        & \\
	         & 1533.4312 & 0.22930 & Noisy  & $~-450$ to $+95~$ & 0.25 & 0.03 & <13.78          &      & 	    &      &        & \\
\ion{Si}{iv}     & 1393.7550 & 0.52800 & 2.7320 & $-1055$ to $+445$ & 3.61 & 0.73 & >15.06\,^{\dag} &	   & >15.06 &      &        & \\
	         & 1402.7700 & 0.26200 & 2.7322 & $-1055$ to $+445$ & 2.79 & 0.37 & >15.06\,^{\dag} &	   &	    &      &        & \\
\ion{S}{ii}      & 1250.5840 & 0.00545 & 2.7333 & $~-340$ to $+15~$ & 0.24 & 0.03 &  15.56          & 0.10 &  15.43 & 0.10 &  -0.28 & 0.14 \\ 
	         & 1253.8110 & 0.01088 & 2.7328 & $~-340$ to $+15~$ & 0.24 & 0.03 &  15.26          & 0.10 &	    &	   &        & \\
\ion{Al}{ii}     & 1670.7874 & 1.88000 & 2.7317 & $~-795$ to $+225$ & 3.19 & 0.45 & >14.12\,^{\dag} &      & >14.12 &	   & >-0.82 & \\
\ion{Al}{iii}    & 1854.7164 & 0.53900 & 2.7323 & $~-415$ to $+30~$ & 1.81 & 0.40 &  14.36          & 0.22 &  14.38 & 0.18 &        & \\
	         & 1862.7896 & 0.26800 & 2.7326 & $~-415$ to $+30~$ & 1.19 & 0.18 &  14.39          & 0.15 &	    &	   &        & \\
\ion{Fe}{ii}     & 1608.4510 & 0.05850 & 2.7313 & $~-795$ to $+150$ & 1.28 & 0.13 &  15.11          & 0.09 &  15.14 & 0.11 &  -0.88 & 0.15 \\
	         & 2344.2129 & 0.11420 & 2.7317 & $~-795$ to $+150$ & 3.16 & 0.41 & >14.97\,^{\dag} &      & 	    &	   &        & \\
	         & 2374.4612 & 0.03131 & 2.7326 & $~-450$ to $+15~$ & 1.46 & 0.21 &  15.17          & 0.13 & 	    &	   &        & \\
	         & 2382.7649 & 0.32000 & 2.7320 & $~-795$ to $+225$ & 4.16 & 1.25 & >14.76\,^{\dag} &      & 	    &	   &        & \\
\ion{Zn}{ii}     & 2026.1360 & 0.48900 & 2.7331 & $~-340$ to $+15~$ & 0.47 & 0.05 &  13.48          & 0.09 & <13.48 &      & <+0.31 & \\
	         & 2062.6641 & 0.25600 & B      & $~-340$ to $+15~$ & 0.29 & 0.03 & <13.53          &	   &	    &      &        & \\
\ion{Cr}{ii}     & 2056.2539 & 0.10500 & ND     & $~-340$ to $+15~$ & 0.48 & 0.05 & <14.16          &	   & <14.09 &      & <-0.12 & \\ 
	         & 2062.2339 & 0.07800 & ND     & $~-340$ to $+15~$ & 0.32 & 0.03 & <14.09          &	   &        &      &        & \\
\ion{Ni}{ii}     & 1317.2170 & 0.05700 & ND     & $~-340$ to $+15~$ & 0.16 & 0.02 & <14.32          &	   & <14.32 &      & <-0.48 & \\
	         & 1741.5531 & 0.04270 & ND     & $~-340$ to $+15~$ & 0.41 & 0.05 & <14.62          &	   &        &      &        & \\
\\
\hline
\end{tabular}
\begin{minipage}{205mm}
\smallskip
\textit{Notes}.---``B'' indicates cases where a reliable value $z_{\rm abs}$ 
could not be measured because of strong blending. ``ND'' refers to cases 
where the line is not detected and only an upper limit to the ion column 
density could be derived. $^{\dag}$ refers to saturated lines where only a 
lower limit to the ion column density could be determined. \\
$^{\rm a}$ Vacuum wavelengths. \\
$^{\rm b}$ Vacuum heliocentric and values measured from the centroid wavelength 
of the interstellar absorption lines. \\
$^{\rm c}$ Velocity range for equivalent width and column density measurements 
relative to $z_{\rm sys} = 2.7350$. \\
$^{\rm d}$ Rest-frame equivalent width and $1\,\sigma$ error. \\
$^{\rm e}$ Adopted ion column density and $1\,\sigma$ error. \\
$^{\rm f}$ Elemental abundance [X/H] = $\log {\rm (X/H)} - 
\log {\rm (X/H)}_{\odot}$ and $1\,\sigma$ error as derived from the adopted ion 
column density relative to the solar meteoritic abundance scale from 
\citet{grevesse07}. 
\end{minipage}
\end{sidewaystable*}
%

%

%

\begin{figure*}
\centering
\includegraphics[width=15cm,clip]{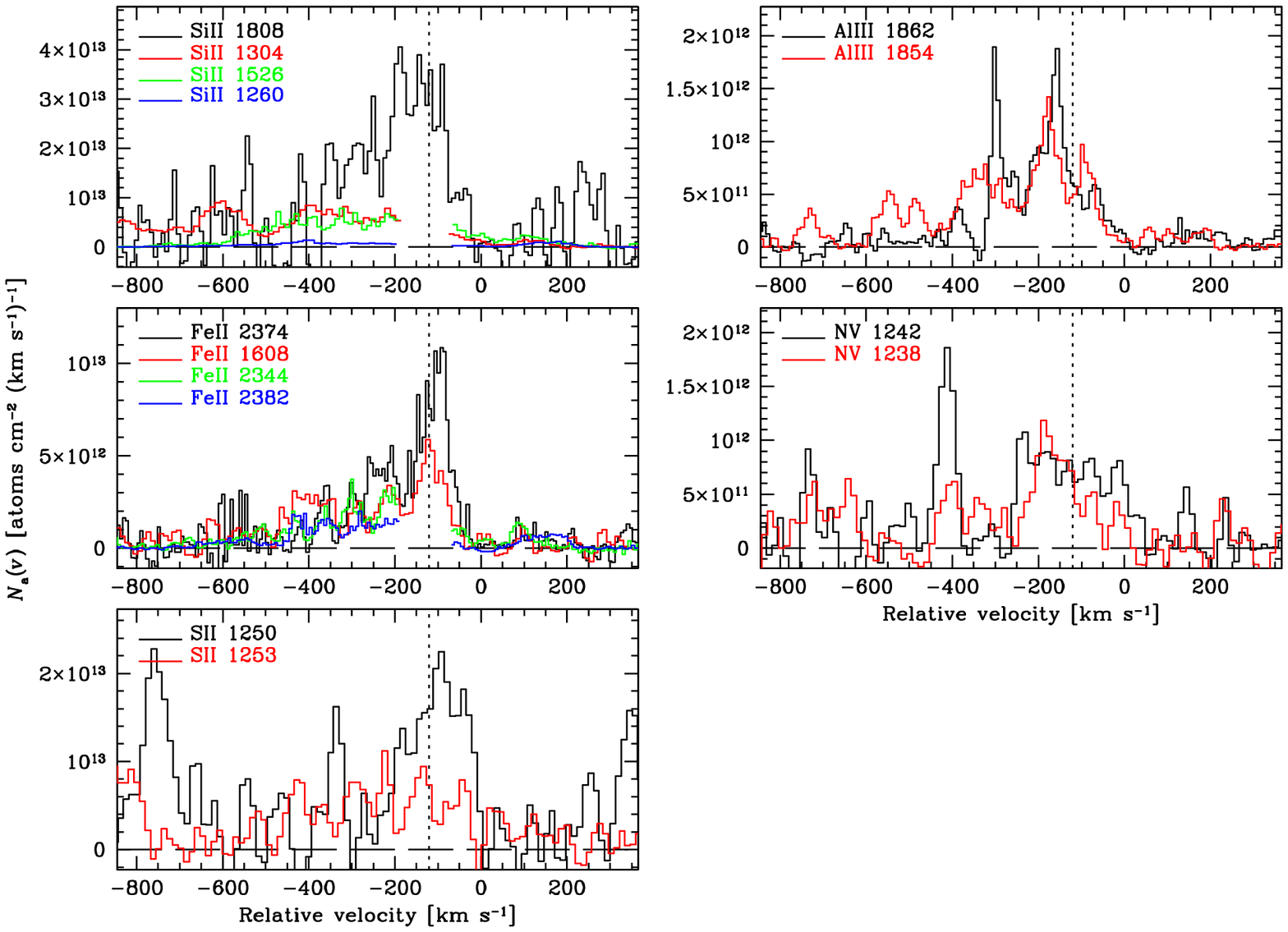}
\caption{Run of apparent column density, $N_{\rm a}(\upsilon)$, as a function
of velocity relative to $z_{\rm sys} = 2.7350$ for absorption lines of various 
ions detected in the 8~o'clock arc. The transitions of a given ion are shown in 
increasing order of $f \lambda_{\rm lab}$, starting from the top. The vertical 
dotted lines at $\upsilon_{\rm ISM} = -120$ km~s$^{-1}$ indicate the redshift 
of ISM gas at the peak optical depth, $z_{\rm ISM}$.}
\label{fig:Na}
\end{figure*}
%

The ISM lines in the ESI spectra of two other well-studied lensed LBGs, the 
Cosmic Horseshoe and MS\,1512--cB58 analyzed by \citet{quider09a} and 
\citet{pettini02}, respectively, have very comparable characteristics 
(broadness, velocity range relative to $z_{\rm sys}$, etc) to the 8~o'clock arc 
lines. The ISM lines of the Cosmic Eye \citep{quider09b} have, on the other
hand, more complex velocity profiles with two well separated components.
\citet{quider09a,quider09b}, moreover, pointed out one noticeable difference 
between the Cosmic Horseshoe plus the Cosmic Eye and cB58 in the optical depth 
of the ISM lines. Whereas in cB58 the strongest ISM lines have saturated cores 
of zero residual intensity (at the resolution of the ESI spectra which is 
comparable to the one of the X-shooter spectra), the same transitions in the 
Horseshoe and Eye seem to reach a minimum residual intensity 
$I_{\rm obs}/I_{\rm c} \sim 0.4$ and $\sim 0.3$, respectively, where 
$I_{\rm obs}$ and $I_{\rm c}$ denote the relative intensities in the line and 
in the continuum, respectively. Quider et~al.\ attributed this effect to the 
fact that the interstellar gas does not completely cover the O and B stars 
producing the UV continuum against which the absorption is seen. The 8~o'clock 
arc, similarly to cB58, does \textit{not} show evidence of partial coverage, 
the strongest ISM lines reach the zero residual intensity in their cores
($I_{\rm obs}/I_{\rm c} \ll 0.1$; see Figs.~\ref{fig:lowions} and 
\ref{fig:highions}). 

%

\subsection{Ion column densities}
\label{sect:columndensities}

Values of column density, $N_{\rm aod}$, for ions listed in 
Table~\ref{tab:ISMlines} were derived using the apparent optical depth method 
of \citet{savage91}. This method is applicable in the case of X-shooter 
spectra despite their intermediate resolution of 42 km~s$^{-1}$ FWHM at 6000 
\AA, because the profiles of the ISM absorption lines in the 8~o'clock arc seem 
to be fully resolved. 

The apparent column density of an ion in each velocity bin, 
$N_{\rm a}(\upsilon)$ in units of atoms~cm$^{-2}$~(km~s$^{-1}$)$^{-1}$, is 
related to the apparent optical depth in that bin, $\tau_{\rm a}(\upsilon)$, by 
the expression
\begin{equation}
N_{\rm a}(\upsilon) = 
\frac{m_{\rm e} c}{\pi {\rm e}^2} \frac{\tau_{\rm a}(\upsilon)}{f \lambda_{\rm lab}} = 
3.768\times 10^{14} \frac{\tau_{\rm a}(\upsilon)}{f \lambda_{\rm lab} (\AA)}\,,
\label{eq:Na}
\end{equation}
where $f$ is the oscillator strength of the transition at the wavelength 
$\lambda_{\rm lab}$ in \AA. The apparent optical depth is deduced directly from 
the observed intensity in the line at velocity $\upsilon$, 
$I_{\rm obs}(\upsilon)$, by
\begin{equation}
\tau_{\rm a}(\upsilon) = -\ln\,[I_{\rm obs}(\upsilon)/I_{\rm c}(\upsilon)]\,,
\label{eq:tau}
\end{equation}
where $I_{\rm c}$ is the intensity in the continuum. The total column density 
of an ion X, $N_{\rm aod}({\rm X})$, is obtained by summation of 
eq.~(\ref{eq:Na}) over the velocity interval where the line absorption takes
place.

The apparent optical depth method provides, in addition, a stringent 
consistency check when several ISM lines arising from the same ground state of 
an ion but with different values of the product $f \lambda_{\rm lab}$ are 
analyzed. The run of $N_{\rm a}(\upsilon)$ with $\upsilon$ should be the same 
for all such lines. In general, this condition will not be satisfied if there 
are \textit{saturated} components in the absorption lines: the deduced value of 
$N_{\rm a}(\upsilon)$ will appear smaller for lines with higher values of 
$f \lambda_{\rm lab}$. A similar effect can also appear if the covering of the 
integrated stellar continuum by the interstellar absorbing gas is inhomogeneous 
at a given velocity. The apparent optical depth method will yield discordant 
values of column density for different transitions of the same ion at that 
velocity.

%

\begin{figure*}
\centering
\includegraphics[width=10cm,clip]{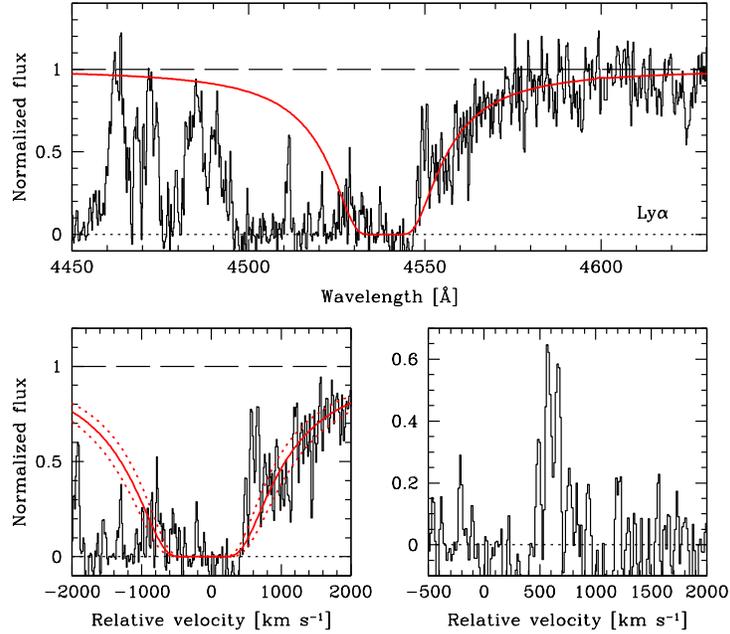}
\caption{\textit{Top panel:} Normalized profile of the Ly$\alpha$ line in the 
8~o'clock arc with the best Voigt profile fit (red solid line) obtained with 
$N({\rm H\,\mathsc{i}}) = 3.7\times 10^{20}$ cm$^{-2}$. 
\textit{Bottom left-hand panel:} Same as top panel, but showing a zoom of the 
Ly$\alpha$ profile in velocity space with the zero velocity fixed to 
$z_{\rm sys} = 2.7350$. Also shown is the $1\,\sigma$ error (red dotted line) 
on the fit of the damped Ly$\alpha$ profile with $N({\rm H\,\mathsc{i}}) = 
2.9\times 10^{20}$ and $4.5\times 10^{20}$ cm$^{-2}$.
\textit{Bottom right-hand panel:} Residual Ly$\alpha$ emission after 
subtraction of the $N({\rm H\,\mathsc{i}}) = 3.7\times 10^{20}$ cm$^{-2}$ 
damped absorption. The zero velocity is still fixed to $z_{\rm sys} =2.7350$.}
\label{fig:Lya}
\end{figure*}
%

In Fig.~\ref{fig:Na} we show the run of $N_{\rm a}(\upsilon)$ with $\upsilon$
for lines of interest. The four \ion{Si}{ii} lines detected have the largest
dynamical range in $f \lambda_{\rm lab}$ values of a factor of 320 from the 
weakest \ion{Si}{ii}\,$\lambda$1808 to the strongest 
\ion{Si}{ii}\,$\lambda$1260 transition. It can be seen from the plots that the 
absorption in the \ion{Si}{ii} lines at velocities between about $-500$ to 
$-100$ km~s$^{-1}$ does not satisfy the consistency check discussed above, in 
that $N_{\rm a}(\upsilon)$ decreases with increasing $f \lambda_{\rm lab}$. The 
effect is particularly dramatic in the core of the line profile centered at 
$\upsilon \simeq -120$ km~s$^{-1}$. This could be indication of either 
saturated absorption components or inhomogeneous coverage of the stellar 
continuum. Saturation is undoubtably at play in \ion{Si}{ii}\,$\lambda$1260, 
\ion{Si}{ii}\,$\lambda$1304, and \ion{Si}{ii}\,$\lambda$1526, because all the
strongest ISM lines reach the zero residual intensity at these velocities (see
Figs.~\ref{fig:lowions} and \ref{fig:highions}). This is further supported by 
the \ion{Fe}{ii} lines. Indeed, with a range in $f \lambda_{\rm lab}$ of a 
factor of 10 from \ion{Fe}{ii}\,$\lambda$2374 to \ion{Fe}{ii}\,$\lambda$2382, 
all the \ion{Fe}{ii} lines satisfy relatively well the consistency check over 
the whole line profile, except in the line core, leaving not much place for 
partial coverage. The discrepancy in the line core between 
\ion{Fe}{ii}\,$\lambda$2374 and \ion{Fe}{ii}\,$\lambda$1608, two lines 
differing by only a factor of 1.28 in their $f \lambda_{\rm lab}$ values, is 
already presumably indicative of saturation in \ion{Fe}{ii}\,$\lambda$1608; and 
this similarly for the discrepancy between the two \ion{S}{ii} lines.

Finally, in the right panels of Fig.~\ref{fig:Na}, we reproduce the run 
$N_{\rm a}(\upsilon)$ with $\upsilon$ for the lines of two high-ions, 
\ion{Al}{iii} and \ion{N}{v}. For both ions we observe a very good consistency 
between their respective line transitions with $f \lambda_{\rm lab}$ values 
differing by a factor of 2. Hence, neither the saturation nor the partial 
coverage seem to affect the line profiles of the high-ions. 

Table~\ref{tab:ISMlines} lists the values of ion column density, $N_{\rm aod}$, 
with the $1\,\sigma$ error, determined by integrating eq.~(\ref{eq:Na}) over 
the velocity interval, $\Delta \upsilon$, spanned by the respective absorption 
lines (same velocity intervals as those used for the equivalent width 
measurements). The lower limits refer to the saturated lines, and the values 
reported as upper limits mainly refer to non-detections. The saturation of a
line is diagnosed via the consistency check discussed above. $N_{\rm adopt}$ 
gives the \textit{adopted} column density for a given ion. It corresponds 
either to the mean of $N_{\rm aod}$ measurements obtained from unsaturated 
lines of a given ion, or to the most stringent limit in case of saturated lines 
and non-detections. We have, nevertheless, to keep in mind that even for ions 
for which we cover at least one transition that is sufficiently weak for the 
apparent optical depth method to be applicable, the derived column density 
values may be \textit{underestimated}, because of possible saturation in the 
line core and possible inhomogeneous coverage of the integrated stellar UV 
light (and saturation) in the line wings.

%

\section{The Ly$\alpha$ line} 
\label{sect:Lyalpha}

\subsection{The damped Ly$\alpha$ profile}
\label{sect:Lyalpha-profile}

The Ly$\alpha$ line profile in the 8~o'clock LBG is a combination of absorption
and emission. In Fig.~\ref{fig:Lya} we show the Ly$\alpha$ line profile and our
decomposition of this feature. The absorption component is best fitted by a
damped profile. We used the software \texttt{FITLYMAN} in \texttt{MIDAS}
\citep{fontana95} which generates theoretical Voigt profiles and performs 
$\chi^2$ minimizations to fit the Ly$\alpha$ profile. The damping wings (and 
in particular the red wing, the blue wing being less constraining and more 
noisy) are well fitted with a neutral hydrogen column density 
$N({\rm H\,\mathsc{i}}) = (3.7\pm 0.8)\times 10^{20}$ cm$^{-2}$ centered at 
$\upsilon_{\rm ISM} \simeq -120$ km~s$^{-1}$, the velocity where the ISM lines 
have the largest optical depth. The derived \ion{H}{i} column density is 
typical of values observed in damped Ly$\alpha$ absorption line systems 
\citep{prochaska05,noterdaeme09}. With the other three \ion{H}{i} column 
densities, $N({\rm H\,\mathsc{i}}) = (7.0\pm 1.5)\times 10^{20}$, 
$(2.5\pm 1.0)\times 10^{21}$, and $(3.0\pm 0.8)\times 10^{21}$ cm$^{-2}$ 
measured in the lensed LBGs MS\,1512--cB58 \citep{pettini02}, FOR\,J0332--3557 
\citep{cabanac08}, and the Cosmic Eye \citep{quider09b}, respectively, and the 
\ion{H}{i} column density measurements for 11 LBGs obtained by
\citet{verhamme08} with the help of their 3D Ly$\alpha$ radiation transfer 
code, the Lyman Break galaxies exhibit a large range of 
$N({\rm H\,\mathsc{i}})$ values, with no trend to a particularly extreme 
reservoir of neutral gas. 

Subtraction of the fit of the damped Ly$\alpha$ absorption reveals a weak 
Ly$\alpha$ emission line (bottom right-hand panel in Fig.~\ref{fig:Lya}). The 
emission is the strongest near $\upsilon_{\rm Ly\alpha} \simeq +570$ 
km~s$^{-1}$, i.e. \textit{redshifted} relative to the systemic redshift of the 
galaxy. It exhibits an asymmetric profile with a relatively abrupt drop on the 
blue side and a more gradual decrease at $\upsilon \gtrsim +700$ km~s$^{-1}$ 
extending up to $\upsilon \sim +1100$ km~s$^{-1}$. The rest-frame equivalent 
width of this Ly$\alpha$ emission line, integrated over the velocity interval 
$\upsilon = +400$ to +1100 km~s$^{-1}$, is $W_0({\rm Ly}\alpha) = -0.35$ \AA. 
However, as radiation transfer models show (see 
Sect.~\ref{sect:radiation-transfer}), this emission peak traces only a very 
small fraction of the intrinsic Ly$\alpha$ emission of the source. 
Table~\ref{tab:summary-z} summarizes the relative velocity measurements 
obtained for the various spectral features in the 8~o'clock arc.

%

\begin{table}
\caption{Relative velocities in the 8~o'clock arc.} 
\label{tab:summary-z}      
\centering  
\begin{tabular}{l   D{=}{\,=\,}{6}   D{.}{.}{1}}   
\hline\hline           
Spectral features & \multicolumn{1}{c}{$z$\,$^{\rm a}$} & \multicolumn{1}{c}{$\upsilon$\,$^{\rm b}$} \\
                  & \multicolumn{1}{c}{}                & \multicolumn{1}{c}{(km~s$^{-1}$)} \\   
\hline                     
Stellar photospheric absorption lines & z_{\rm stars}  	            = 2.7347 & -24 \\
\ion{H}{ii} emission lines            & z_{\rm C\,\mathsc{iii}]}    = 2.7350 & 0 \\
Recombination emission line           & z_{{\rm Si\,\mathsc{ii}}^*} = 2.7352 & +16 \\
Interstellar absorption lines         & z_{\rm ISM}		    = 2.7335 & -120 \\
Ly$\alpha$ emission line              & z_{\rm Ly\alpha}	    = 2.7420 & +570 
\\
\hline                       
\end{tabular}
\begin{minipage}{81mm}
\smallskip
$^{\rm a}$ Vacuum heliocentric. \\
$^{\rm b}$ Relative to the systemic redshift $z_{\rm sys} = 2.7350$ that is the 
mean of $z_{\rm stars}$ and $z_{\rm em}$ (the mean redshift of emission lines).
\end{minipage}
\end{table}
%

The Ly$\alpha$ emission profile in the 8~o'clock arc is remarkably similar to 
that of MS\,1512--cB58 \citep{pettini00,pettini02}, except that the emission in 
the 8~o'clock arc is more redshifted than in cB58 by $>200$ km~s$^{-1}$. The 
lensed LBG studied by \citet{cabanac08} exhibits an even more redshifted 
emission peaking at about +720 km~s$^{-1}$. Redshifted Ly$\alpha$ emission is 
often seen in high-redshift galaxies and in local \ion{H}{ii} and starburst 
galaxies. This redshift results from large-scale outflows of the interstellar 
media. Indeed, Ly$\alpha$ emission is suppressed by resonant scattering and the 
only Ly$\alpha$ photons that can escape unabsorbed in the observer's direction 
are those backscattered from the far side of the expanding nebula, whereas in 
absorption against the stellar continuum, we see the approaching part of the 
outflow. The velocity offset between the Ly$\alpha$ emission and low-ionization 
ISM lines measured in the 8~o'clock arc is typical of $z\sim 3$ LBGs in 
general, and agrees well with the offset observed for LBGs with the strongest 
Ly$\alpha$ absorption \citep{shapley03}.

%

\subsection{Radiation transfer modeling}
\label{sect:radiation-transfer}

\subsubsection{MCLya code and input parameters}

To model the Ly$\alpha$ line of the 8~o'clock arc, we use an improved version 
of the Monte Carlo radiation transfer code, MCLya, of \citet{verhamme06} 
including the detailed physics of Ly$\alpha$ line and UV continuum transfer, 
dust scattering, and dust absorption for arbitrary 3D geometries and velocity 
fields. The following improvements have been included: angular redistribution 
functions taking quantum mechanical results for Ly$\alpha$ into account 
\citep{stenflo80,dijkstra08}, frequency changes of Ly$\alpha$ photons due to 
the recoil effect \citep[e.g.,][]{zheng02}, the presence of deuterium 
\citep[assuming a canonical abundance of ${\rm D/H} = 3\times 10^{-5}$,][]
{dijkstra06}, and anisotropic dust scattering using the Henyey-Greenstein 
phase function (with parameters as adopted in \citet{witt00}). Furthermore, a 
relatively minor bug in the angular redistribution of Ly$\alpha$ photons has 
been fixed, and the code has been parallelized for efficient use on 
supercomputers. For the physical conditions used in the simulations of the 
present paper, these improvements lead only to minor changes with respect to 
the MCLya version used by \citet{schaerer08} and \citet{verhamme08}. More 
details on the code upgrades will be given in Hayes et~al.\ (2009, in
preparation).

For simplicity and as in earlier modeling of $z \sim 3$ LBGs \citep{verhamme08}, 
we assume a simple geometry, i.e.\ a spherical, homogeneous shell of cold ISM 
(neutral hydrogen plus dust) surrounding the starburst (UV continuum plus 
Ly$\alpha$ line emission from the \ion{H}{ii} region). The input parameters of 
the 3D transfer simulations are: the radial expansion velocity, \vexp, the 
\ion{H}{i} column density, $N({\rm H\,\mathsc{i}})$, the \ion{H}{i} velocity 
dispersion, $b$, and the dust absorption optical depth, \taua, which expresses 
the dust-to-gas ratio. As discussed in \citet{verhamme06}, \taua\ is related to 
the usual color excess $E(B-V)$ by $E(B-V) \approx (0.06...0.11) \taua$. We 
assume $E(B-V) = 0.1 \taua$ for convenience. 

For each parameter set, a full Monte Carlo (MC) simulation is run. As described 
in \citet{verhamme06}, our MC simulations are computed for a flat input 
spectrum, keeping track of the necessary information to recompute \textit{a 
posteriori} simulations for arbitrary input spectra. For the Ly$\alpha$ fits, 
we assume an input spectrum given by a flat (stellar) continuum plus the 
Ly$\alpha$ line described by a Gaussian with variable equivalent width, 
$W({\rm Ly}\alpha$), and full width at half maximum, FWHM(Ly$\alpha$). In a 
first step, we use an automatic line profile fitting tool, relying on an 
extensive grid of 5200 MC simulations and exploring the full parameter space 
(Hayes et~al.\ 2009, in preparation). Some of the model parameters (e.g., 
\vexp) are fixed in a subsequent step to take observational constraints into 
account. Shortward of 4528 \AA\ the blue wing of the Ly$\alpha$ line is 
affected by other lines (see Fig.~\ref{fig:Lya}) and is therefore excluded from 
our line fitting procedure.

%

\begin{figure}
\centering
\includegraphics[width=8.8cm,clip]{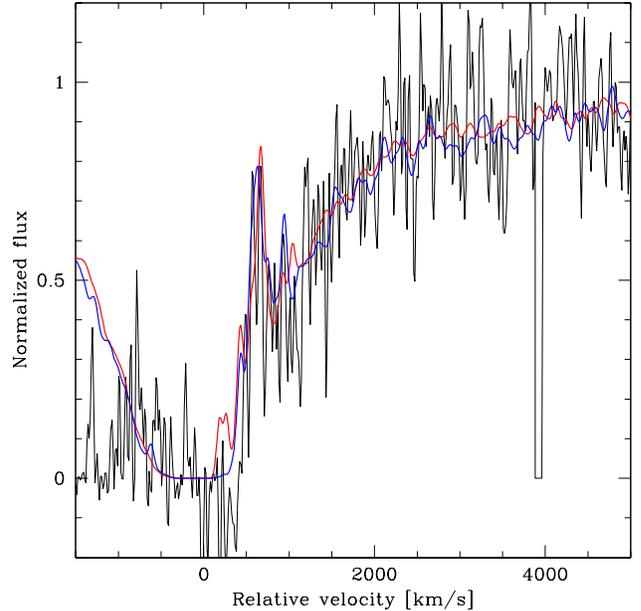}
\caption{Ly$\alpha$ line profile fits (red/blue) to the observed profile 
(black) of the 8~o'clock arc using a modified version of the 3D Ly$\alpha$ and 
dust radiation transfer code of \citet{verhamme06} applied to homogeneous, 
spherically expanding shells surrounding a UV continuum + line source. The 
redshift and the radial outflow velocity are fixed at $z = z_{\rm sys}$ and 
$\vexp = 150$ km~s$^{-1}$, respectively, as determined from the observations. 
The remaining fit parameters are $\log N({\rm H\,\mathsc{i}}) = 20.8$ (20.8)
cm$^{-2}$, $\taua = 3$ (4), $b = 20$ (80) km~s$^{-1}$, $W({\rm Ly}\alpha) = 
-50$ ($-50$) \AA, and ${\rm FWHM(Ly}\alpha) = 50$ (160) km~s$^{-1}$ for the red 
(blue) curves, respectively.}
\label{fig:Lyafit}
\end{figure}
%

\subsubsection{Results}

Our automated Ly$\alpha$ fitting code yields the following best-fit parameters
for the Ly$\alpha$ line profile of the 8~o'clock arc: $\vexp = 0^{+52}_{-0}$
km~s$^{-1}$, $\log N({\rm H\,\mathsc{i}}) = 20.8^{+0.6}_{-0.03}$ cm$^{-2}$, 
$\taua = 3^{+0.96}_{-2.8}$, $b = 40^{+46}_{-30}$ km~s$^{-1}$, 
$W({\rm Ly}\alpha) = -300^{+299}_{-0}$ \AA, and ${\rm FWHM(Ly}\alpha) = 
100^{+0}_{-49}$ km~s$^{-1}$, where the errors correspond to the formal 
1-dimensional 68\% confidence levels. Here, negative equivalent widths indicate 
emission. Although these formal errors are quite large, one should notice that 
strong correlations exist between various parameters. For example, reducing the 
dust content (\taua) requires a higher \ion{H}{i} column density to maintain 
the same width of the broad Ly$\alpha$ absorption.

The most interesting result is the finding of a high \ion{H}{i} column density 
and a large dust optical depth, which (for the closed geometry adopted here) 
are necessary to create a damped Ly$\alpha$ absorption line, as already shown 
in \citet{verhamme06} and \citet{schaerer08}. The derived \ion{H}{i} column 
density exceeds the one determined in Sect.~\ref{sect:Lyalpha-profile} from a 
simple Voigt profile fitting by a factor of $\sim 1.7$, again this is expected 
given the assumption on the geometry of a shell surrounding the UV source 
\citep{verhamme06}. This difference in the $N({\rm H\,\mathsc{i}})$ estimates 
(although geometry dependent) has implications on the abundance determinations 
discussed in Sect.~\ref{sect:metallicities}. The best-fit value of \taua\ 
corresponds to an UV attenuation $E(B-V) \approx 0.3$, in reasonable agreement 
with the color excess $E(B-V) = 0.67\pm 0.21$ derived from the Balmer decrement 
by \citet{finkelstein09}.

Despite an overall good fit of the damped Ly$\alpha$ absorption component in 
the 8~o'clock arc, the above solution fails to reproduce the Ly$\alpha$ 
emission component observed in the red wing of the Ly$\alpha$ absorption (see
Sect.~\ref{sect:Lyalpha-profile} and Fig.~\ref{fig:Lya}). However, this 
emission is easily recovered when assuming larger expansion velocities. In 
particular, for $\vexp \sim 100-150$ km~s$^{-1}$, an emission peak appears 
naturally in the red wing and at the observed wavelength. Two examples of 
such fits are shown in Fig.~\ref{fig:Lyafit}. This bulk radial velocity of the 
shell is in agreement with the mean outflow velocity of $-120$ km~s$^{-1}$ 
measured from the low-ionization interstellar absorption lines relative to the 
systemic redshift (see Sect.~\ref{sect:ISMlines} and 
Table~\ref{tab:summary-z}). For higher velocities, the peak is shifted too far 
to the red. The input, i.e.\ intrinsic FWHM of the Ly$\alpha$ emission line 
before undergoing radiation transfer, does not much affect the result. Adopting 
e.g.\ ${\rm FWHM(Ly}\alpha) = 160$ km~s$^{-1}$, as suggested by the 
\ion{C}{iii}] lines, does not alter the resulting profile. The results are more 
sensitive to $b$, which describes the adopted velocity dispersion in the 
expanding shell. Good fits are obtained with $b \la 90$ km~s$^{-1}$, as also 
shown in Fig.~\ref{fig:Lyafit}. Larger values of $b$ lead to a broader
Ly$\alpha$ emission peak, when strong intrinsic Ly$\alpha$ emission is present. 
These values for $b$, corresponding to ${\rm FWHM} \la 210$ km~s$^{-1}$, are 
reasonable, as can be judged from the column density weighted velocity 
distribution of the low-ionization ISM lines (see Fig.~\ref{fig:Na}). From this 
we conclude that our radiation transfer models, assuming a simple expanding 
shell and ISM properties in good agreement with observations, are able to 
reproduce the observed Ly$\alpha$ profile, including the broad absorption and 
the small peak of ``reminiscent'' Ly$\alpha$ emission. More sophisticated 
models would require a knowledge of the presumably more complex ISM geometry 
and velocity field.

The Ly$\alpha$ profile fits of the 8~o'clock arc shown in Fig.~\ref{fig:Lyafit} 
correspond to an input \textit{intrinsic} Ly$\alpha$ emission with a rest-frame 
equivalent width of $W_0({\rm Ly}\alpha) = -50$ \AA, a typical value expected 
when star formation extends over timescales longer than $\ga 10^7$ yr. The 
Ly$\alpha$ fit for this LBG therefore agrees well with our earlier findings for 
other LBGs \citep{schaerer08,verhamme08}. In particular, our radiation transfer 
models show that the Ly$\alpha$ line profile of the 8~o'clock arc is compatible 
with an approximately constant star formation, where the intrinsic Ly$\alpha$ 
emission is transformed into the complex, observed profile by radiation 
transfer effects and absorption by dust.

The Ly$\alpha$ emission peak in the red wing of the Ly$\alpha$ absorption is 
the result of multiply backscattered Ly$\alpha$ line photons emitted in the 
\ion{H}{ii} region surrounded by the cold, expanding shell, as explained in 
\citet{schaerer08} for cB58. The velocity comparison of the observed Ly$\alpha$ 
emission peak ($500-700$ km~s$^{-1}$) to the outflow velocity ($120$ 
km~s$^{-1}$) in the 8~o'clock arc (see Table~\ref{tab:summary-z}) indicates 
that these emergent Ly$\alpha$ photons have benefited from multiple ($2-3$) 
backscattering across the shell. This relatively large number of 
backscattering is due to the high \ion{H}{i} column density. It explains why 
the velocity shift of the Ly$\alpha$ emission peak is found at 
$\upsilon_{{\rm Ly}\alpha}/\vexp \sim 4-6$, larger than the ``typical'' shift 
of $\upsilon_{{\rm Ly}\alpha}/\vexp \sim 2$ suggested by \citet{verhamme06} for 
LBGs with lower $N({\rm H\,\mathsc{i}})$.

The observed velocity offset between the Ly$\alpha$ emission peak and the 
low-ionization ISM lines of $\Delta \upsilon({\rm Ly}\alpha - {\rm ISM}) \sim 
620-820$ km~s$^{-1}$ in the 8~o'clock arc is also in good agreement with the 
average $\Delta \upsilon({\rm Ly}\alpha - {\rm ISM}) \sim 790$ km~s$^{-1}$ 
measured from the composite spectra of $z\sim 3$ LBGs with strong Ly$\alpha$ 
absorption \citep[see][]{shapley03}. For cB58, in contrast, $\Delta 
\upsilon({\rm Ly}\alpha - {\rm ISM}) \sim 550$ km~s$^{-1}$. In this respect, 
the 8~o'clock arc appears to be more typical of the category of LBGs with 
damped Ly$\alpha$ profiles, and our modeling results for this LBG therefore 
support the explanation of \citet{verhamme08} for the observed correlation 
between $\Delta \upsilon({\rm Ly}\alpha - {\rm ISM})$ and $W_0({\rm Ly}\alpha)$ 
shown by \citet{shapley03}. In short, variations of the velocity offset between 
the Ly$\alpha$ emission and the ISM absorption lines are strongly affected by 
the interstellar medium column density and do not primarily reflect changes in 
outflow velocities.

%

\begin{table}
\caption{Metallicity estimates in the 8~o'clock arc.} 
\label{tab:summary-metallicity}      
\centering  
\begin{tabular}{l c l c}   
\hline\hline           
Environment & Method & Elements & Z/Z$_{\odot}$ \\
\hline
\ion{H}{ii} regions$\,^{\rm a}$ & N2                     & O         & 0.83 \\
\ion{H}{ii} regions$\,^{\rm a}$ & O3N2                   & O         & 0.39 \\
OB stars\,$^{\rm b}$            & ``1425'' index         & C, Si, Fe & 0.85 \\
OB stars\,$^{\rm b}$            & ``1978'' index         & Fe        & 0.79 \\
ISM gas\,$^{\rm c}$             & Apparent optical depth & Si        & 0.65
\\
\hline                       
\end{tabular}
\begin{minipage}{73mm}
\smallskip
$^{\rm a}$ As derived by \citet{finkelstein09}. \\
$^{\rm b}$ This work (Sect.~\ref{sect:stellarlines}). \\
$^{\rm c}$ This work (Sect.~\ref{sect:columndensities}). 
\end{minipage}
\end{table}
%

\section{Summary of the results and discussion}
\label{sect:discussion}

\subsection{Consistency in the various metallicity estimates}
\label{sect:metallicities}

Lensed Lyman Break galaxies observed with current instrumentation are the only 
objects at high redshifts where metallicity estimates from stars, \ion{H}{ii} 
regions, and interstellar gas are all accessible. They thus offer a nice 
comparison of these various environments. In Table~\ref{tab:summary-metallicity} 
we summarize all the available metallicity estimates in the 8~o'clock arc.

We determine the metallicity of OB stars from the photospheric absorption 
lines and the related ``1370'', ``1425'', and ``1978'' metallicity indices 
(Sect.~\ref{sect:stellarlines}). The well-calibrated ``1425'' and ``1978'' 
indices lead to metallicities, $Z=0.85~Z_{\odot}$ and $Z=0.79~Z_{\odot}$, 
respectively, which are in very good mutual agreement. \citet{finkelstein09} 
obtained the oxygen abundance of the ionized gas from nebular emission lines 
redshifted into the NIR, using the N2 and O3N2 indices of \citet{pettini04}. 
The derived O3N2 metallicity is only half of the N2 metallicity. The authors 
assigned this discrepancy to the uncertainty in their [\ion{O}{iii}] flux 
measurement, and consider the N2 metallicity, $Z=0.83~Z_{\odot}$, as being the 
more reliable. The metallicity of early-type stars thus is in excellent 
agreement with that of the \ion{H}{ii} regions that surround the stars. We 
naturally expect the two metallicities to be the same, since the stars 
presumably formed very recently out of the gas which they now ionize. The 
corresponding mean metallicity is $Z=0.82~Z_{\odot}$.

The 8~o'clock arc with this metallicity and its stellar mass estimate of 
$\sim 4.2\times 10^{11}$~M$_{\odot}$ \citep{finkelstein09} is consistent with 
the mass-metallicity relation at $z\sim 2.2$ of 
\citet{erb06b}\footnote{Established on the same metallicity calibration index 
N2 as the one used by \citet{finkelstein09} for their metallicity estimate.}, 
although it lies at a slightly higher mass than the highest mass points of Erb 
et~al. We may have expected the 8~o'clock arc to fall below the $z\sim 2.2$ 
trend, given its higher redshift and the observed trend in the mass-metallicity 
relation to move downward in metallicity objects from low-to-high redshifts 
\citep{maiolino08}. However, this is not the case, as also stated by
\citet{finkelstein09}.

We also determine the metallicity in the interstellar medium of the 8~o'clock 
arc from the numerous ISM absorption lines detected 
(Sect.~\ref{sect:columndensities}). The $Z=0.65~Z_{\odot}$ metallicity of the
ISM gas, as determined from the silicon abundance, is about 80\% the 
metallicity of OB stars and ionized gas, i.e.\ only $\sim 0.1$~dex lower. When 
taking these measurements at face value, they suggest that the interstellar 
medium of the 8~o'clock LBG has rapidly been polluted by ejecta from OB stars 
and enriched to the metallicity of \ion{H}{ii} regions. On the other hand, 
several uncertainties affect these measurements: (i)~the radiation transfer 
modeling seems to show that the \ion{H}{i} column density is underestimated by 
a factor of $\sim 1.7$ when derived from a simple Voigt profile fitting of the 
damped Ly$\alpha$ profile (see Sect.~\ref{sect:radiation-transfer}), if this is 
correct the ISM gas metallicity would also be a factor of $\sim 1.7$ lower than 
the stellar and ionized gas metallicities; (ii)~without the possibility to 
carry out a photoionization analysis, we assume that 
${\rm Si\,\mathsc{ii}}/{\rm H\,\mathsc{i}} = {\rm Si/H}$, but there may be 
some need for ionization corrections; and (iii)~some saturation in the line 
cores and a possible inhomogeneous coverage of the stellar light in the line 
wings may lead to an underestimation of the metal column densities. How 
significant these effects are, remains difficult to quantify with our data. 
Metallicity differences are observed between the ionized gas and the neutral 
ISM gas in some nearby dwarf galaxies \citep[see e.g.,][]{aloisi03,
lebouteiller09}, but even these local examples are of controversial 
interpretation.

%

\subsection{Elemental abundances in the interstellar medium}

From the ion column densities (Sect.~\ref{sect:columndensities}) and the 
\ion{H}{i} column density derived from the damped Ly$\alpha$ profile 
(Sect.~\ref{sect:Lyalpha-profile}), we determine the chemical abundances of 
several elements in the interstellar medium of the 8~o'clock arc. They are 
listed, with their $1\,\sigma$ errors, in the last two columns in 
Table~\ref{tab:ISMlines} relative to the solar meteoritic abundance scale from 
\citet{grevesse07}. 

Our X-shooter observations cover three $\alpha$-capture elements, Si, S, and O. 
Si and S give a consistent picture within measurement uncertainties, with 
abundances ${\rm [Si/H]} = -0.19\pm 0.14$ and ${\rm [S/H]} = -0.28\pm 0.14$, 
respectively. The lower limit on the abundance of O is, on the other hand, 
useless, given the strong saturation of the \ion{O}{i}\,$\lambda$1302 line and 
its blend with \ion{Si}{ii}\,$\lambda$1304. 

As for the iron-peak elements, we have the abundance measurement of Fe and the 
upper limits on the abundances of Zn, Cr, and Ni. The Fe abundance, 
${\rm [Fe/H]} = -0.88\pm 0.15$, is lower than that of $\alpha$-elements by 
0.69~dex (Si). This underabundance could be a departure from the solar scale 
due to nucleosynthetic effects, a reflection of depletion of Fe onto dust 
grains, or both. The abundance of Zn, an element that is undepleted, usually 
helps to break the above ambiguity \citep[e.g.,][]{pettini99}. Unfortunately, 
our super-solar Zn abundance, ${\rm [Zn/H]} < +0.31$ (super-solar abundances 
are not observed for any other elements), favor that the detection of the 
\ion{Zn}{ii}\,$\lambda$2026 line is extremely marginal (see 
Fig.~\ref{fig:lowions}) and that the derived Zn abundance is only a 
non-constraining upper limit.

The gas-phase abundance of the interstellar Fe does not also agree with the Fe 
abundance in the OB stars as deduced from the ``1978'' index which arises from 
the blending of numerous \ion{Fe}{iii} transitions 
(see Table~\ref{tab:summary-metallicity}) and which does not suffer from dust 
depletion. At face value, ${\rm (Fe/H)}_{\rm stars} \simeq 6\times 
{\rm (Fe/H)}_{\rm ISM}$. This could be an indication that most of the 
underabundance of the interstellar Fe in the 8~o'clock arc is due to dust 
depletion. If dust depletion is the sole origin of the Fe underabundance 
relative to the $\alpha$-elements, then, according to what is observed in the 
interstellar medium of the Milky Way \citep{savage96}, Si should also be 
depleted. This is, however, not what is observed, as the abundance of Si is 
approximately the same to the abundance of the undepleted S, as well as to the 
abundance of O as derived from \ion{H}{ii} regions and to the abundance in the 
OB stars (see Table~\ref{tab:summary-metallicity}). Nevertheless, the 
conditions in the ISM of these actively star-forming galaxies, that are the 
LBGs, are likely to be quite different from those in the Milky Way, where the 
star formation rate is about 100 times lower.
We are hence unable to quantify the respective contributions from dust 
depletion and nucleosynthesis to the underabundance of Fe relative to the
$\alpha$-elements. An intrinsic overabundance of the $\alpha$-capture products 
of Type~II supernovae relative to the iron-peak elements whose release into the 
ISM is delayed, because produced on much longer timescales by Type~Ia 
supernovae, would suggest a relatively \textit{young} ($\la 0.3-1$ Gyr) age for 
the bulk of stars in the 8~o'clock LBG. Such an age limit is in agreement with 
the spectral energy distribution of the 8~o'clock arc 
\citep[see][]{finkelstein09} and with the strength of the 
\ion{He}{ii}\,$\lambda$1640 emission line (see Sect.~\ref{sect:emissionlines}).

%

\subsection{Comparison with other Lyman Break galaxies}

The lensed 8~o'clock arc offers a new opportunity to compare the detailed 
properties of individual Lyman Break galaxies and establish how typical are the
properties of MS\,1512--cB58, the first lensed LBG studied \citep{pettini00,
pettini02,teplitz00}. \citet{quider09a,quider09b} provided the analysis of the 
second and the third lensed LBGs, the Cosmic Horseshoe and the Cosmic Eye,
respectively, made at a comparable precision thanks to their 
intermediate-resolution ESI spectra. \citet{quider09a} discuss the similarities 
and differences between the Horseshoe and cB58 in light of $z\sim 2-3$ 
star-forming galaxies. Here we would like to add the new example of the 
8~o'clock arc.

\subsubsection*{General characteristics}

After correction for lensing, with $\sim 11~L^*$ the 8~o'clock arc is the most 
luminous Lyman break galaxy relative to cB58, the Horseshoe, and the Eye in the 
rest-frame UV \citep{allam07}. But, the lensed 8~o'clock LBG is not only among 
the most luminous LBGs. Dynamical masses of $M_{\rm dyn} \sim 1-1.7\times 
10^{10}$~M$_{\odot}$, typical of LBGs, are measured for cB58, the Horseshoe, 
and the Eye \citep[see][]{teplitz00,coppin07,hainline09}. In comparison, the 
stellar mass of the 8~o'clock arc is estimated to $M_{\rm stars} \sim 4.2\times 
10^{11}$~M$_{\odot}$ \citep{finkelstein09}, namely significantly more massive. 
cB58 has the lower extinction $E(B-V) \sim 0.27$ compared to the Horseshoe and 
the 8~o'clock arc with $E(B-V)\sim 0.45$ and $\sim 0.67$, respectively, as 
derived from the Balmer decrement. The extinction corrected H$\alpha$ and 
H$\beta$ star formation rates of cB58, the Horseshoe, and the Eye are 
${\rm SFR} \sim 50-100$ M$_{\odot}$~yr$^{-1}$ \citep{stark08,quider09a}. The 
star formation rate of the 8~o'clock arc, ${\rm SFR} \sim 270$ 
M$_{\odot}$~yr$^{-1}$, is significantly higher, namely higher than $\sim 85$\% 
of star-forming galaxies at $z\sim 2-3$ \citep{finkelstein09}.

%

\subsubsection*{The metallicity}

The metallicities ($Z\simeq 0.4-0.5~Z_{\odot}$) of cB58, the Horseshoe, and the
Eye are very comparable, whereas the metallicity ($Z\sim 0.8~Z_{\odot}$) of the 
8~o'clock arc is sensibly higher. This is in-line with the mass-metallicity 
relation derived from samples of $z\sim 2.2$ and $z\sim 3.5$ star-forming 
galaxies \citep{erb06b,maiolino08}. Indeed, given the $\sim 11~L^*$ luminosity 
of the 8~o'clock arc and its high stellar mass, we expect it to have a 
metallicity higher than that for typical $L^*$ galaxies at this redshift. The 
8~o'clock LBG, with its high metallicity and high mass, is located at the upper 
end of the LBG mass-metallicity distribution, and may hence appear as less 
representative of the whole LBG population. All the Lyman Break galaxies at 
$1<z<3$, with their metallicities in the range of $\sim 0.3$ to $0.9~Z_{\odot}$ 
seem to have already achieved a near-solar metallicity at relatively early 
times, some 12 Gyr ago in the case of the 8~o'clock arc, when the age of the 
Universe was only 17\% of what it is today. This advanced degree of chemical 
enrichment is consistent with the original suggestion by \citet{steidel96} that 
LBGs are the progenitors of today's ellipticals and bulges, since such 
relatively high abundances are common in the most massive galaxies at 
$z\simeq 3$. 

%

\begin{figure}
\centering
\includegraphics[width=7.5cm,clip]{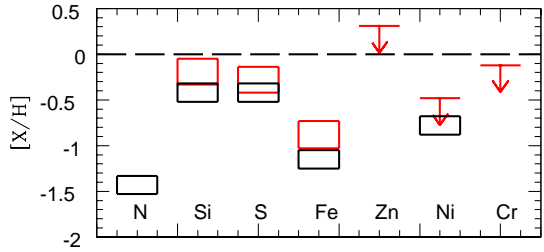}
\caption{Comparison of the elemental abundances, [X/H], in the interstellar 
medium of the 8~o'clock arc (red) and MS\,1512--cB58 (black). The height of the 
boxes reflects the $1\,\sigma$ error on the abundance measurement, and the 
arrows refer to upper limits.}
\label{fig:abundances}
\end{figure}
%

\subsubsection*{The interstellar medium}

cB58, the Horseshoe, and the 8~o'clock arc show remarkable similarities in the
kinematic properties of their interstellar medium gas: (i)~first of all the 
blueshift of the ISM lines relative to the stars attributed to large-scale 
outflows, also observed in the large samples of $z\sim 2-3$ star-forming 
galaxies; (ii)~the large broadness of the ISM lines with velocities spanning 
$\sim 1000$ km~s$^{-1}$, from about $-800$ to $+300$ km~s$^{-1}$ relative to 
the systemic redshift; and (iii)~the similarity of the ISM line profiles among 
all ion species. As for the chemical abundances, the lensed LBGs seem also to 
present common characteristics. In Fig.~\ref{fig:abundances} we compare the 
elemental abundances measured in cB58 and the 8~o'clock arc. They both show a 
very nice agreement between the $\alpha$-capture abundances, and a relatively 
high $\alpha$-enhancement relative to Fe-peak elements ($\sim 0.7$~dex) which
may be attributed to both dust depletion and/or nucleosynthesis. 
Galaxies associated with damped Ly$\alpha$ absorption line systems (DLAs) also 
offer the opportunity to study in detail the chemical abundances in the 
interstellar medium at high redshifts. The pattern of elemental abundances in 
LBGs appears to be different from DLAs, as already pointed out by 
\citet{pettini02}. First, as for the metallicity, DLAs at all redshifts are 
generally metal-poor, the probability to find one DLA with a metallicity 
$Z>0.4~Z_{\odot}$ at $z=2-3$ is lower than 1/10 \citep[e.g., the recent 
compilation by][]{dessauges09}. Second, the enhancement of $\alpha$-elements 
relative to Fe-peak elements due to nucleosynthetic effects seems very elusive 
to pin down in these galaxies \citep{prochaska02,dessauges06}. These chemical 
differences suggest that star formation does not proceed in the same way in 
LBGs as in DLAs \citep{jimenez99}. There are also obvious differences in the 
kinematics of the ISM gas between LBGs and DLAs. The major one is the broadness 
of the ISM line profiles that span only $\sim 300$ km~s$^{-1}$ in DLAs 
\citep[e.g.,][]{wolfe00a,wolfe00b}. Furthermore, contrary to LBGs, in DLAs 
significant differences are observed between the profiles of low- and 
high-ionization lines. Given all these distinctions, DLAs very likely trace 
another galaxy population at high redshift than the LBGs.

\subsubsection*{The \ion{He}{ii} emission}

The detection of the \ion{He}{ii}\,$\lambda$1640 emission line in the 8~o'clock 
arc and in FOR\,J0332--3557 studied by \citet{cabanac08} betrays the presence 
of very massive stars---the Wolf-Rayet stars---in these high-redshif galaxies. 
In nearby galaxies, this and other WR features are well known tracers of 
massive star-forming regions \citep{conti91,schaerer98,schaerer99}. In distant 
starbursts with high star formation rates, the timescale of their star 
formation activity is expected to be most likely relatively long compared to 
the lifetime of the massive stars ($\la 10$ Myr). When star formation proceeds 
over long timescales, recently updated models predict a \ion{He}{ii} emission 
with equivalent widths $W_0({\rm He\,\mathsc{ii}}) \sim -0.5$ \AA\ for 
metallicities $\sim 0.4$ solar, and larger for higher metallicities 
\citep{brinchmann08,schaerer98}. The observations of the few lensed LBGs appear 
in agreement with this prediction. Indeed, the 8~o'clock arc with its high 
metallicity, $Z \sim 0.8~Z_{\odot}$, shows a strong \ion{He}{ii}\,$\lambda$1640 
line. FOR\,J0332-3557 also shows a strong \ion{He}{ii} emission, although no 
accurate metallicity estimate is currently available for this LBG. In contrast, 
in cB58 and the Horseshoe, both characterized by lower metallicities, $Z\sim 
0.5~Z_{\odot}$, the \ion{He}{ii} emission has not been detected, being quite 
likely below the detection threshold. Alternatively, differences in their star 
formation histories, age dependent extinction \citep{leitherer02}, or other 
effects could be invoked to explain the difference in strength of the 
\ion{He}{ii} line. Finally, it is also natural that 
$W_0({\rm He\,\mathsc{ii}})$ is larger in the 8~o'clock arc than in the 
composite spectrum of $z\sim 3$ LBGs, as our object is among the brightest, 
most massive, and most metal-rich LBGs.

\subsubsection*{The Ly$\alpha$ profile}

The most striking difference between LBGs and also lensed LBGs observed so far 
is certainly the morphology of their Ly$\alpha$ profile. While in cB58, the 
8~o'clock arc, FOR\,J0332--3557, and the Eye the Ly$\alpha$ line is dominated 
by a damped absorption profile on top of which is superposed a weak emission
(except in the Eye, where no emission is observed), in the Horseshoe the 
Ly$\alpha$ line is characterized by a strong double-peak emission profile. The 
common property of the emission component is its redshift relative to the ISM 
absorption lines, also observed in the large samples of $z\sim 2-3$ 
star-forming galaxies. With its broad Ly$\alpha$ absorption profile and the 
available detailed information, the 8~o'clock arc offers a new opportunity to 
test the scenario proposed by \citet{schaerer08} and \citet{verhamme08} to 
explain the Ly$\alpha$ absorption in LBGs and the diversity of other observed 
Ly$\alpha$ line profiles. In fact, it turns out that our radiation transfer 
models work better for the 8~o'clock arc than for cB58 analyzed previously, 
where we had to account for deviations from a spherical shell model 
\citep{schaerer08}. A homogeneous spherical shell model with a constant outflow 
velocity, determined by the observations, is able in the case of the 8~o'clock 
arc to reproduce the observed Ly$\alpha$ line profile, and the required dust 
content agrees well with the attenuation measured from the Balmer decrement. 
Furthermore, the assumption of homogeneity is reasonable in the 8~o'clock arc, 
since the low-ionization ISM absorption lines indicate a (nearly) complete 
coverage of the UV source. The results obtained from the fit of the Ly$\alpha$ 
line therefore fully support the scenario proposed earlier \citep{schaerer08,
verhamme08}, where we showed that the diversity of Ly$\alpha$ line profiles in 
LBGs and Ly$\alpha$ emitters (LAEs), from absorption to emission, is mostly due 
to variations of \ion{H}{i} column density and dust content. This scenario also 
naturally explains the main correlations observed between the Ly$\alpha$ 
emission and other properties of LBGs highlighted by \citet{shapley03}. In 
particular, our detailed spectrum of the 8~o'clock arc and the fit of the 
Ly$\alpha$ line support the explanation for the observed correlation of the 
velocity shift between the Ly$\alpha$ emission and ISM lines, 
$\Delta \upsilon({\rm Ly}\alpha - {\rm ISM})$, with $W_0({\rm Ly}\alpha)$. The 
increase of $\Delta \upsilon({\rm Ly}\alpha - {\rm ISM})$ with increasing 
Ly$\alpha$ absorption is mostly due to an increase of the \ion{H}{i} column 
density, as nicely observed between the 8~o'clock arc and 
FOR\,J0332--3557\footnote{In the 8~o'clock arc we measure $\Delta 
\upsilon({\rm Ly}\alpha - {\rm ISM}) \sim 690$ km~s$^{-1}$ for 
$\log N({\rm H\,\mathsc{i}}) = 20.57$, while in FOR\,J0332--3557 we measure 
$\Delta \upsilon({\rm Ly}\alpha - {\rm ISM}) \sim 830$ km~s$^{-1}$ for 
$\log N({\rm H\,\mathsc{i}}) = 21.40$.}, which implies that multiple 
scattering/radiation transfer effects become more important. Despite the 
relatively high values of $\Delta \upsilon({\rm Ly}\alpha - {\rm ISM})$, the 
bulk outflow velocities remain relatively modest in galaxies with strong 
Ly$\alpha$ absorption. 
Measuring the ISM absorption and the stellar photospheric absorption line 
redshifts remains the most reliable method to determine outflow velocities.

\subsubsection*{UV covering factor and ISM geometry}

The UV covering factor of the cold ISM gas in LBGs has been determined for 
cB58, the Horseshoe, and the Eye. While in cB58 the blackness of the strongest 
interstellar absorption lines indicates a nearly complete coverage of the UV 
continuum by the ISM \citep{pettini02}, strong evidence for a patchy ISM with a 
coverage of only $\sim 60$\% and $70-85$\% of the UV continuum was found in the 
Horseshoe and the Eye, respectively \citep{quider09a,quider09b}. Our data of 
the 8~o'clock arc show no evidence for a partial coverage, resembling again the 
case of cB58. What determines these differences and which case may be more 
general for LBGs remains to be determined. Independently of this question, it 
is clear that the partial UV coverage must also affect the Ly$\alpha$ line 
profile, leading preferentially to strong Ly$\alpha$ emission, as discussed by 
\citet{quider09a}. In the Cosmic Eye, the unusual presence of several, 
especially redshifted, components of cold gas, the large extinction 
\citep{smail07}, and the exceptionally high \ion{H}{i} column density may very 
well explain the absence of any Ly$\alpha$ emission. If partial UV coverage 
was a common phenomenon, the unifying scenario explaining the diversity of 
Ly$\alpha$ line profiles in LBGs and LAEs with changes in \ion{H}{i} column 
density and dust content (see above) could not be upheld. Clearly, further 
detailed studies on the ISM of high-redshift galaxies are needed to address 
this and other related questions thoroughly.

%

\begin{acknowledgements}
The good quality of the spectra obtained in the first nights of the instrument 
at the telescope is the result of the successful efforts of the X-shooter 
consortium team. More than 60 engineers, technicians, and astronomers worked 
over more than 5 years on the project in Denmark, France, Italy, the 
Netherlands, and at ESO. We recall here in representation of the whole team the
co-PIs P.~Kjaergaard-Rasmussen, F.~Hammer, R.~Pallavicini, L.~Kaper, and 
S.~Randich. R.~Pallavicini, one of the strongest supporter of the project, died 
prematurely just after the first light of the instrument. Special thanks go 
also to the ESO Commissioning team led by H.~Dekker and including among others 
J.~Lizon, R.~Castillo, M.~Downing, G.~Finger, G.~Fischer, C.~Lucuix, E.~Mason, 
and P.~Santin. We thank the anonymous referee for a helpful and constructive 
report, and we are also grateful to M.~Pettini for his useful comments. MDZ and 
DS are supported by the Swiss National Science Foundation.
\end{acknowledgements}

%

\end{document}